%% file: sn-article.tex
\documentclass[pdflatex]{sn-jnl}

\usepackage{natbib}

\usepackage{graphicx}%
\usepackage{multirow}%
\usepackage{amsmath,amssymb,amsfonts}%
\usepackage{amsthm}%
\usepackage{mathrsfs}%
\usepackage[title]{appendix}%
\usepackage{xcolor}%
\usepackage{textcomp}%
\usepackage{manyfoot}%
\usepackage{geometry}
\usepackage{algorithm}%
\usepackage{algorithmicx}%
\usepackage{algpseudocode}%
\usepackage{listings}%
\usepackage{bm}
\input{header}

\begin{document}

\title[Compresisble MHD turbulence]{\textbf{Magnetized compressible turbulence with a fluctuation dynamo and Reynolds numbers over a million}}


\author*[1,2,3]{\fnm{James} \sur{R. Beattie}}\email{james.beattie@princeton.edu}
\author[1,4]{\fnm{Christoph} \sur{Federrath}}
\author[5,6]{\fnm{Ralf} \sur{Klessen}}
\author[7]{\fnm{Salvatore} \sur{Cielo}}
\author[2]{\fnm{Amitava} \sur{Bhattacharjee}}

\affil[1]{\orgdiv{Research School of Astronomy and Astrophysics}, \orgname{Australian National University}, \orgaddress{\city{Canberra}, \postcode{2611}, \state{ACT}, \country{Australia}}}

\affil[2]{\orgdiv{Department of Astrophysical Sciences}, \orgname{Princeton University}, \orgaddress{\city{Princeton}, \postcode{08540}, \state{NJ}, \country{USA}}}

\affil[3]{\orgdiv{Canadian Institute for Theoretical Astrophysics}, \orgname{University of Toronto}, \orgaddress{\city{Toronto}, \postcode{M5S3H8}, \state{ON}, \country{Canada}}} 
 
\affil[4]{\orgname{Australian Research Council Center of Excellence in All Sky Astrophysics (ASTRO3D)}, \orgaddress{\city{Canberra}, \postcode{2611}, \state{ACT}, \country{Australia}}} 

\affil[5]{\orgdiv{Zentrum f\"ur Astronomie, Institut f\"ur Theoretische Astrophysik}, \orgname{Universit\"at Heidelberg}, \orgaddress{\city{Heidelberg}, \postcode{69120}, \state{Baden-W\"urttemberg}, \country{Germany}}} 

\affil[6]{\orgdiv{Interdisziplin\"ares Zentrum f\"ur Wissenschaftliches Rechnen}, \orgname{Universit\"at Heidelberg}, \orgaddress{\city{Heidelberg}, \postcode{69120}, \state{Baden-W\"urttemberg}, \country{Germany}}} 

\affil[7]{\orgname{Leibniz Supercomputing Centre of the Bavarian Academy of Sciences and Humanities}, \orgaddress{\city{Garching}, \postcode{85748}, \state{Bavaria}, \country{Germany}}} 


\abstract{

\textbf{Supersonic magnetohydrodynamic (MHD) turbulence is a ubiquitous state for many astrophysical plasmas. However, even the basic statistics for this type of turbulence remains uncertain. We present results from supersonic MHD turbulence simulations at unparalleled resolutions, with plasma Reynolds numbers of over a million. In the kinetic energy spectrum we find a break between the scales that are dominated by kinetic energy, with spectral index $-$2, and those that become strongly magnetized, with spectral index $-$3/2. By analyzing the Helmholtz decomposed kinetic energy spectrum, we find that the compressible modes are not passively mixed through the cascade of the incompressible modes. At high magnetic Reynolds number, above $10^5$, we find a power law in the magnetic energy spectrum with spectral index $-$9/5. On the strongly magnetized, subsonic scales the plasma tends to self-organize into locally relaxed regions, where there is strong alignment between the current density, magnetic field, velocity field and vorticity field, depleting both the nonlinearities and magnetic terms in the MHD equations, which we attribute to plasma relaxation on scales where the magnetic fluctuations evolve on shorter timescales than the velocity fluctuations. This process constrains the cascade to inhomogenous, volume-poor, fractal surfaces between relaxed regions, which has significant repercussions for understanding the nature of magnetized turbulence in astrophysical plasmas and the saturation of the fluctuation dynamo.}

}

\maketitle

\section{Introduction}\label{sec:introduction}
    Magnetohydrodynamic (MHD) turbulence is ubiquitous across many scales in the Universe, whether it be the solar wind, accretions disks, each of the phases of the interstellar medium (ISM) of galaxies, or even the intracluster medium (ICM) between galaxies \citep{MacLow2004,Bruno2013_soloar_wind_turbulence,Bruggen2015_ICM_turbulence,Ferriere2020_reynolds_numbers_for_ism,Fielding2022_ISM_plasmoids,Rosotti2023_supersonic_turbulence_in_pp_discs,Hopkins2023_forged_in_fire_II}. In these plasmas, turbulence plays a diverse set of roles, from the transport of magnetic flux, momentum, and thermal energy, to regulating the rate in which cold gas is turned into stars \citep{MacLow2004,Federrath2012,Mohapatra2019_turbulent_heat_flux_ICM,Kempski2022_cr_scattering}. In the cold phases of the ISM, and in thin accretions disks, the rest frame root-mean-squared (rms) turbulent velocity fluctuations $\delta u$ on the outer scale of the turbulence $\ell_0$ are trans-to-supersonic, $\delta u/c_s \gtrsim 1$, where $c_s$ is the local sound speed \citep{Beattie2019b,Federrath2016_brick,Ferriere2020_reynolds_numbers_for_ism,Rosotti2023_supersonic_turbulence_in_pp_discs,Hopkins2023_forged_in_fire_II}. Even in solar wind plasmas, characterized by strong magnetic fields (thermal-to-magnetic pressure ratios --\emph{plasma beta}-- of the order of unity or less) and $\delta u / c_s \ll 1$, the turbulence is weakly compressible and shows significant deviations from incompressible MHD turbulence phenomenology \citep{Zank_1992_waves_in_solar_wind,Bhattacharjee_1998_weakly_compressible_solar_wind}. All of these plasmas harbor extremely large plasma Reynolds numbers that facilitate the nonlinear transfer of energy across vast ranges of scales through the turbulence cascade, and exhibit high levels of magnetization, maintained by a variety of magnetic dynamos \citep{Rincon2019_dynamo_theories,Brandenburg2023_galactic_dynamo_review}. Therefore, to understand the turbulence in these plasmas, one has to understand supersonic, compressible MHD turbulence at extremely high Reynolds numbers. However, most of the theories and large Reynolds number calculations are for incompressible, subsonic turbulence \citep{Iroshnikov_1965_IK_turb,Kraichnan1965_IKturb,Goldreich1995,Boldyrev2006,Beresnyak2014_4k_incomp_sim,Galishnikova2022_saturation_and_tearing,Dong2022_reconnection_mediated_cascade}, limiting their applicability for understanding the basic statistical properties of these supersonic turbulent systems.

    \begin{figure}[]
        \centering
        \includegraphics[width=\linewidth]{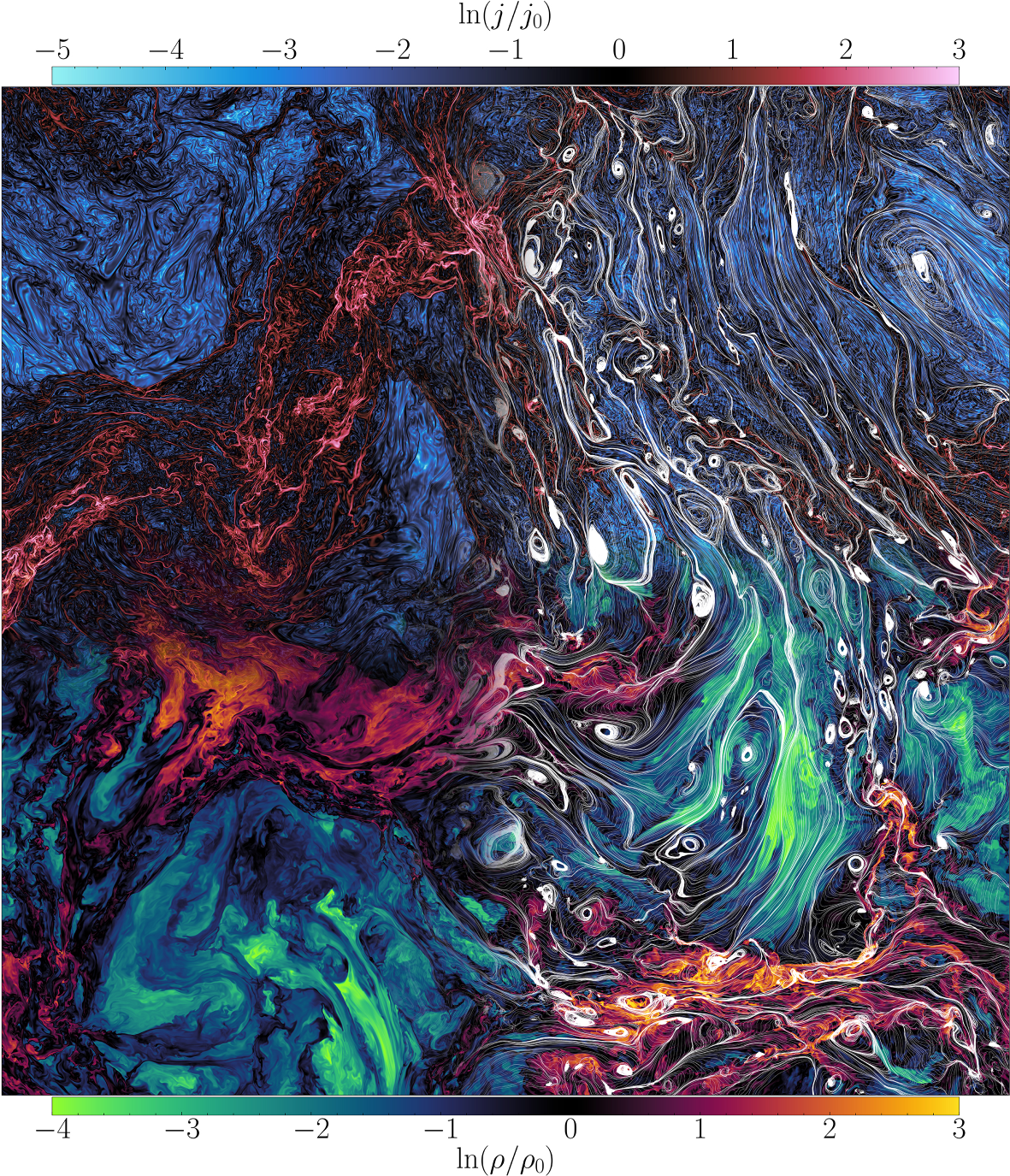}
        \caption{\textbf{The gas density, magnetic field and current density structure in the world's largest supersonic MHD turbulence simulation.} A two-dimensional slice from the $10080^3$ supersonic MHD simulation showing the logarithmic magnitude of the current density $\ln(j/j_0)$ (top), gas density $\ln(\rho/\rho_0)$ (bottom), and magnetic field streamlines (right; white), where the subscript zero indicates the volume average. Fractal current sheets are distributed through the densest regions, accompanied by tightly coiled magnetic fields.}
        \label{fig:slice_poster}
    \end{figure}

    Hydrodynamical, supersonic turbulence has been simulated at hydrodynamical Reynolds numbers $\Re = \delta u \ell_0 / \nu \gtrsim 10^6$ where $\nu$ is the coefficient of kinematic viscosity, demonstrating the existence of a scale-separated double turbulent cascade in the kinetic energy spectra \citep{Federrath2021}. Based on the second-order structure functions, scales $\ell_s < \ell < \ell_0$ exhibit a \citet{Burgers1948} cascade, with $\ekin(k) \sim k^{-2}$, while scales $\ell_\nu < \ell < \ell_s$ follow a \citet{Kolmogorov1941} cascade, with $\ekin(k) \sim k^{-5/3}$ (with \citet{She1994} intermittency corrections), where $k = 2\pi/\ell$ is the wavenumber, $\ekin(k)$ is the kinetic energy spectrum, $\ell_s$ represents the sonic scale where $\delta u(\ell_s) = c_s$ and $\ell_\nu$ is the viscous dissipation scale. No such calculation exists for supersonic MHD turbulence at $\Re \gtrsim 10^6$, until the simulation presented in this study. Analogous to the hydrodynamical Reynolds number, the magnetic Reynolds number is defined as $\Rm = \delta u \ell_0 / \eta$, where $\eta$ is the plasma resistivity.
    Similar to supersonic turbulence, the numerical evidence for a break scale in three-dimensional, incompressible MHD turbulence has been accumulating over only the last few years \citep{Dong2022_reconnection_mediated_cascade,Galishnikova2022_saturation_and_tearing,Fielding2022_ISM_plasmoids,Grete2023_as_a_matter_of_dynamical_range}, precluded by the theory of extremely high $\Rm$ reconnection-driven plasmoid instabilities that disrupt and modify the cascade $\emag(\kperp) \sim \ekin(\kperp) \sim \kperp^{-11/5}$, where $\kperp$ is the wavevector perpendicular to the large-scale magnetic field \citep{Comisso2018_MHD_turbulence_plasmoid_regime,Boldyrev2020_tearing_mode_instability,Dong2022_reconnection_mediated_cascade}. To capture this phenomenon, simulations require $\Rm\gtrsim 10^5$ so that thin current sheets become unstable to the tearing instability \citep{Bhattacharjee2009_fast_reconnection,Uzdensky2010_fast_reconnection,Loureiro_2016_plasmoid_instability,Boldyrev2017_MHD_mediated_by_reconnection,Mallet2017_plasmoid_disruptions}. This happens on scales where the instability growth timescale is shorter than the dynamical timescale of the turbulence, leading to the break scale $k_{*} \sim \Rm^{4/7}$ \citep{Loureiro2017_reconnection_in_turbulence,Dong2022_reconnection_mediated_cascade}. This has been measured in a recent strong-field, decaying MHD turbulence simulation at $\Rm \sim 10^5$ \citep{Dong2022_reconnection_mediated_cascade}, but it is still an important and open question if it will be realized in other turbulence regimes.

    Further pertinent to the study of driven MHD turbulence is the concept of alignment. In strong Alfv\'enic turbulence, the nonlinear timescale $t_{\rm nl}$ associated with the MHD cascade rate is set by interactions between counter-propagating Alfv\'en wavepackets $\mathbf{z}_{\perp}^{\mp} = \bfu_{\perp}\mp\bfb_{\perp}$ that cascade perpendicular to the magnetic field, where $\bfb_{\perp}$ is in velocity units. It is $t_{\rm nl} \sim 1/(k_{\perp}\mathbf{z}^{\mp}_{\perp}\sin\theta)$, where $\theta$ is the angle between $\mathbf{z}^+_{\perp}$ and $\mathbf{z}^{-}_{\perp}$. To change $t_{\rm nl}$ one can change the alignment between $\mathbf{z}^+$ and $\mathbf{z}^{-}$ \citep{Boldyrev2006,Perez2009_dynamical_alignment_of_imbalanced_islands,Mallet2015_refined_cb}, or put the turbulence in a state where $\bfu\propto\pm\bfb$ \citep{Dobrowolny1980_solar_wind_properties,Chernoglazov2021_alignment_SR_MHD}, the latter completely halting the cascade. One can attain such a highly-aligned $\bfu$ and $\bfb$ configuration by realizing a minimum energy state \citep{Matthaeus2008_rapid_alignment}, or a maximum entropy state \citep{Banerjee2023_relaxed_states}, through the process of plasma relaxation. For a relaxation process that approximately preserves the rugged, inviscid, MHD invariants, magnetic helicity and cross helicity, this gives $\bfu \propto \bfb \propto \bfj \propto \bm{\omega}$, (see \ref{app:relaxation}) where $\bfj = (1/\mu_0)\bm{\nabla}\times\bfb$ is the current density, $\mu_0$ is the magnetic permittivity, and $\bm\omega = \bm{\nabla} \times \bfu$ is the fluid vorticity. Hence, such a process does not only tend to align $\bfu$ and $\bfb$, as in dynamical alignment \citep{Boldyrev2006,Perez2009_dynamical_alignment_of_imbalanced_islands}, but requires alignment to be spread also across $\bfu$ and $\bfb$ and the respective curls, in turn depleting all of the nonlinearities of the MHD equations (see \ref{app:relaxation}). Such states may be realized through a number of different processes, are sensitive to the energy flux from the turbulent driving source, and may only be observed if there is significant scale-separation between $\ell_0$ and scales within the cascade \citep{Banerjee2023_relaxed_states}, as is the case for the simulations presented in our study.

    We aim to address the nature of the spectral and alignment statistics of turbulence in a regime that combines both supersonic rms velocities and $\Re \sim \Rm \gtrsim 10^6$ -- an unexplored parameterisation of turbulence -- also requiring that the magnetic field we use is self-consistently generated from a magnetic dynamo (as opposed to an imposed field). This requires an extremely large dynamical range that resolves many orders of magnitude of spatial scales in the turbulence, from the large-scale shocks to the thin current sheets, hence, this can only be done by pushing numerical codes and computing infrastructure to the very limits of modern capacities.

\section{Results}\label{sec:results}
    \subsection{The world's largest supersonic MHD turbulence simulation:}
        We present the first results from an ensemble of driven, supersonic, $\delta u / c_s =  4.3 \pm 0.2$, magnetized turbulence simulations that have a magnetic field being self-consistently maintained by the turbulent dynamo in saturation, providing a volume integral energy ratio of $\emag/\ekin = 0.242 \pm 0.022$. The grids vary from $2520^3$ ($\Rm \sim \Re \sim 10^5)$ up to $10080^3$ ($\Rm \sim \Re \sim 3\times10^6$), discretised on a triply-periodic domain with length $L$. Presently, these are the largest supersonic, magnetized simulations in the world, almost an order of magnitude larger in grid resolution (and Reynolds numbers) compared to previous simulations in this regime \citep{Fielding2022_ISM_plasmoids,Grete2023_as_a_matter_of_dynamical_range} and are the first simulations to resolve both a supersonic and subsonic cascade with a self-consistently maintained magnetic field. The simulations utilised over 80~million CPU hours distributed across nearly 140,000~compute cores on the high-performance supercomputer, SuperMUC-NG, at the Leibniz Supercomputing Centre. We integrate the $10080^3$ simulation for $t \approx 2t_0$, where $t_0 = \ell_0 / \delta u$ is the turnover time on the driving scale of the turbulence $\ell_0 = L/2$ (or equivalently $kL/2\pi = 2$), allowing for time-averaging of all key statistics across $\approx 2t_0$, making for robust, statistically significant results. We provide details on the simulation methods in \ref{sec:methods}.

    \subsection{Structure of the gas and current density:} 
        In Figure~\ref{fig:slice_poster} we show a two-dimensional slice of the $10080^3$ simulation, with the logarithmic, mean-normalized magnitude of the current density $\ln(j/j_0)$ (top) and gas density $\ln(\rho/\rho_0)$ (bottom), vertically blended together and overlaid with magnetic field streamlines on the right half. The zero subscript indicates the mean over the entire volume. In force balance, which shocked gas tends toward, $c_s^2\bm{\nabla} \rho = \bfj \times \bfb/\mu_0$ \citep{Robertson2018,Mocz2018}. Hence, complex interactions of intense sheets of current (shown in red) form along orthogonal surfaces to $\bm{\nabla} \rho$. Close inspection of the sheets in $\ln(j/j_0)$ reveals that plasmoid-like instabilities develop sparsely in the most intense currents spread through the voids (blue), where the shear flow is smallest (shown with a zoom-in Figure~\ref{fig:plasmoids} in \ref{app:plasmoids}). The outer scale of the plasmoids is $k_{*}L/2\pi \gtrsim 10^2$, which are well-resolved in our simulations. The lifetime of the voids can be orders of magnitude longer than the dense structures, which exist for a fraction of a sound crossing time \citep{Hopkins2013_non_lognormal_s_pdf,Robertson2018,Mocz2019,Beattie2022_spdf}. Thus it is possible that the supersonic motions significantly disrupt the small-scale current sheets in the dense regions, and the plasmoids are only able to form in the current sheets embedded in the longer lived voids, where the shear flow is the weakest. If these instabilities are constrained to sparsely populate the voids, this may prevent them from influencing the global statistics in this regime, as in \citep{Dong2022_reconnection_mediated_cascade}. The $\ln(\rho/\rho_0)$ field shows shocked, gaseous filaments in red, and deep voids in green, with fluctuations $-4 \leq \ln(\rho/\rho_0) \leq 3$, highlighting how the gas density varies by orders of magnitude, typical of supersonic turbulence \citep{Federrath2010_solendoidal_versus_compressive,Beattie2022_spdf}.

    \begin{figure}
        \centering
        \includegraphics[width=\linewidth]{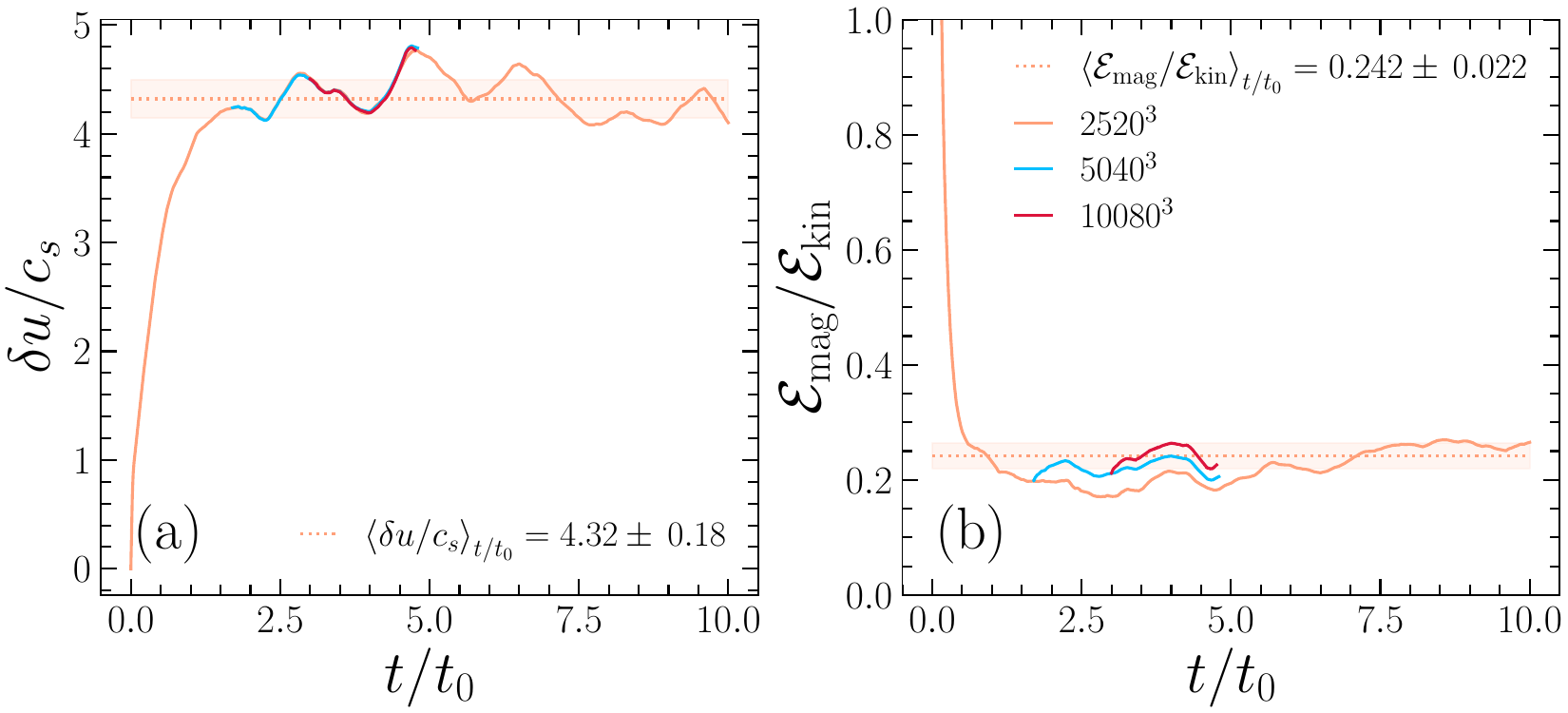}
        \caption{\textbf{The time evolution of the volume integral root-mean-squared velocity fluctuations and magnetic to kinetic energy ratios.} \textbf{(a):} The root-mean-squared velocity fluctuations in units of sound speed $\delta u / c_s$ as a function of time in units of correlation times $t_0$ for the $2520^3$ (orange), $5040^3$ (blue) and $10080^3$ (red) simulations. Each higher-resolution simulation is carried out from an initial condition of a linearly interpolated version of the lower-resolution simulation, avoiding the initial non-stationary state, i.e., the $0 \leq t/t_0 \lesssim 2$ range of times for the $2520^3$ simulation. The average and 1$\sigma$ are shown for the last $5t_0$ with the dashed line and bounding box, respectively, giving $\delta u / c_s = 4.32 \pm 0.18$. \textbf{(b):} the same as (a) but for the integral magnetic to kinetic energy ratio, $\emag/\ekin$. In the steady state, $\emag/\ekin = 0.242 \pm 0.022$, as desired (see \ref{app:init_conditions}).}
        \label{fig:integral_quantities}
    \end{figure}

    \subsection{Time evolution of volume integral quantities}\label{app:volume_integral_quants}
\begin{figure}
        \centering
        \includegraphics[width=\linewidth]{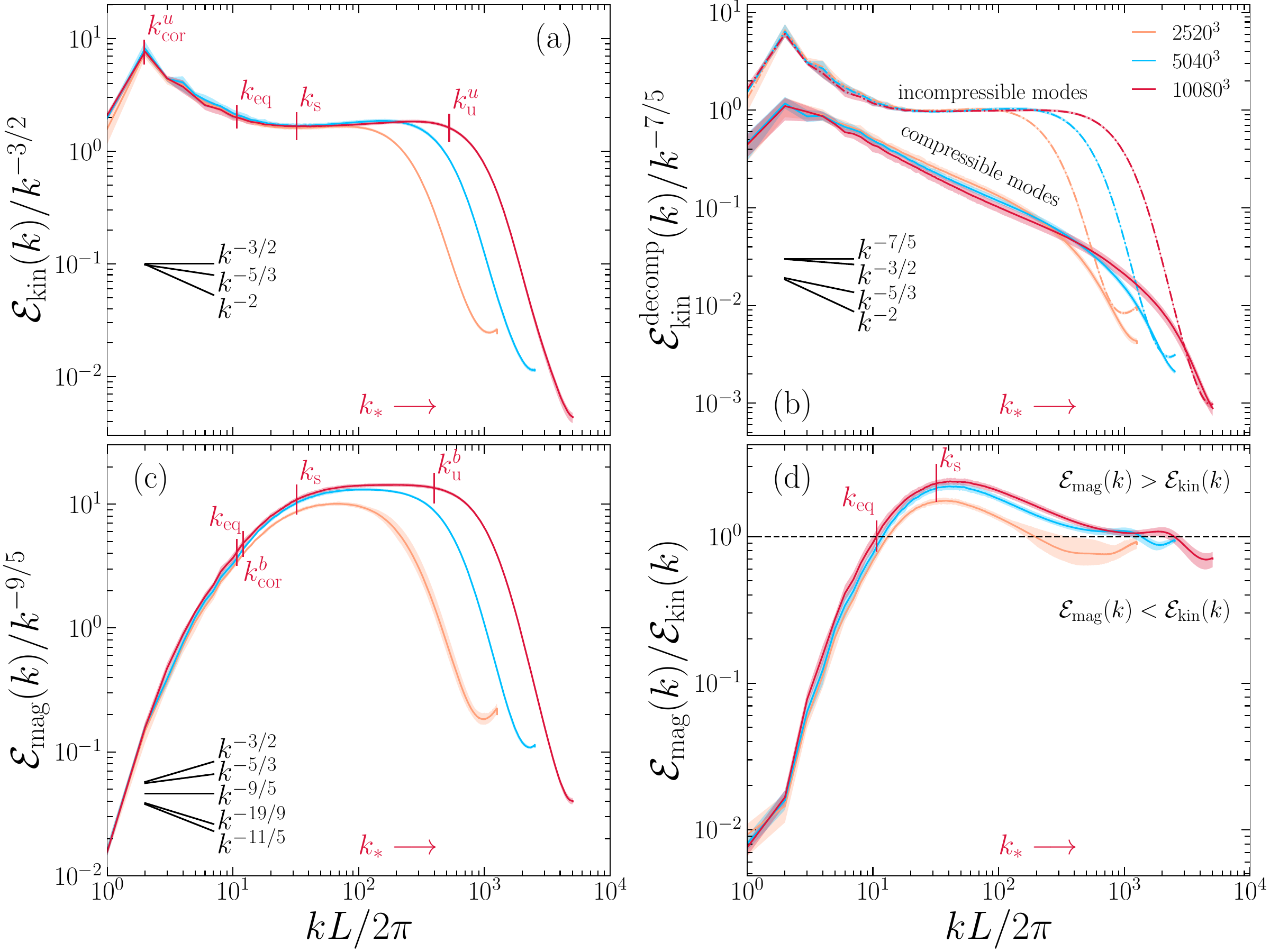}
        \caption{\textbf{The energy spectra and fundamental turbulence scales.} The correlation scale, $\kcor$, energy equipartition scale, $\keq$, sonic scale, $\ks$, inner scale, $\ku$, and upper bound for the plasmoid outer scale, $k_{*}$, are annotated in the panels (see \ref{app:turbulent_scales} and \ref{app:plasmoids} for definitions). Each color corresponds to a different grid resolution. \textbf{(a)}: The kinetic energy spectra $\ekin(k)$ compensated by $k^{-3/2}$. We observe a transition between scales separated by $\keq$, where the turbulence goes from hydrodynamically to magnetically dominated (see panel d). On large scales, $\ekin(k) \sim k^{-2}$, whilst on small scales, $\ekin(k) \sim k^{-3/2}$. \textbf{(b)}: The $\ekin(k)$ decomposed into compressible (solid line) and incompressible (dashed line) modes. Each spectra traces a different slope, suggesting compressible modes are not passive to incompressible modes \citep{Lithwick2001_compressibleMHD}. \textbf{(c)}: The magnetic energy spectrum $\emag(k)$ compensated by $k^{-9/5}$, with the same annotated scales as in (a). The compensation shows an extended power law, $\emag(k) \sim k^{-9/5}$, at $k > \ks$. Different turbulence power law models are shown for comparison: dynamical alignment, $k^{-3/2}$ \citep{Boldyrev2006}, unaligned strong turbulence, $k^{-5/3}$ \citep{Goldreich1995}, an empirical relation, $k^{-9/5}$ \citep{Fielding2022_ISM_plasmoids}, tearing instability in dynamo, $k^{-19/9}$ \citep{Galishnikova2022_saturation_and_tearing}, and the reconnection-mediated cascade, $k^{-11/5}$ \citep{Dong2022_reconnection_mediated_cascade}. \textbf{(d)}: The $\emag(k)/\ekin(k)$ energy ratio showing the $\keq$ transition, and the peak of $\emag(k)/\ekin(k)$ at $\sim\ks$.}
        \label{fig:spectra}
    \end{figure}

        We show the time evolution of $\delta u / c_s$ (a) and $\emag/\ekin$ (b) in and Figure~\ref{fig:integral_quantities}, showing the time-averaged value for each of the quantities in the legend. The different colors represent the different simulation grids and, as discussed in the previous section, from this plot it is easy to observe where in time each of the simulations were started from, using the hierarchical interpolation technique discussed in the previous section. These plots convey that all the simulations are indeed in a stationary state. Between the different grids, the values of $\delta u / c_s = 4.32 \pm 0.18 \approx 4$, shown in panel (a) are nearly perfectly matching over time, likely due to the $\delta u / c_s$ being dominated by the low $k$ velocity modes, so changing the grid spacing, and in effect $\Rm$ and $\Re$, has little effect on the $\delta u / c_s$ statistics. However, there are minor differences in $\emag/\ekin = 0.242 \pm 0.022 \approx 1/4$, shown in panel (b) as we change the simulation resolution, with $\emag/\ekin$ growing as we increase the resolution. This can be attributed to the fact that the magnetic field, and hence $\emag(k)$, is inherently a small-scale, high-$k$ mode-dominated field, in comparison to the velocity field. This feature is well-recognized in the dynamo community \citep{Schekochihin2004_dynamo,Federrath2016_dynamo,Galishnikova2022_saturation_and_tearing,Beattie2023_growth_or_decay}, and we show it explicitly in Figure~\ref{fig:spectra} where we separate the $\emag(k)$ and $\ekin(k)$ spectra.

    \subsection{Energy spectra and fundamental turbulence scales:}
        This unique numerical experiment resolves a broad range of scales and challenges profoundly the tenets of MHD turbulence theories. We show this directly by measuring the time-averaged, isotropic $\ekin(k)$ (a) and $\emag(k)$ (c) in Figure~\ref{fig:spectra} (see \ref{app:turbulent_scales} for spectra definitions). $\ekin(k)$ and $\emag(k)$ are significantly different from one another, both structurally and in amplitude, as shown with the $\emag(k)/\ekin(k)$ ratio, Figure~\ref{fig:spectra}~(d). This deviates significantly from current incompressible MHD turbulence theories, which predict the same power laws for $\emag(k)$ and $\ekin(k)$ \citep{Schekochihin2020_bias_review}. We find a \citet{Burgers1948} spectrum $\ekin(k) \sim k^{-2}$ on large scales, where $\delta u(k) > c_s$, $\kcor^u \sim k_0 > k > \keq = (10.6\pm 0.7) 2\pi/L$, with $\kcor^u$ the correlation scale of the turbulence and $\keq$ the $\ekin(\keq) = \emag(\keq)$ energy equipartition (or MHD) scale. On the $\delta u(k) < c_s$ range of scales $\ks = (32\pm2) 2\pi/L < k < \ku^u$ where $\ks$ is the sonic scale and $\ku^u$ is the inner scale, we find an \citet{Iroshnikov_1965_IK_turb}-\citet{Kraichnan1965_IKturb} (IK) spectrum  $\ekin(k) \sim k^{-3/2}$. The $\emag(k)/\ekin(k)$ spectrum is maximized at $\ks$, hence the sonic transition $\delta u(\ks) = c_s$ is the most magnetized scale in the turbulence. This is a similar dichotomy as found in supersonic, hydrodynamical turbulence \citep{Federrath2021}, but now with the additional effect that the magnetic field is dominant at $k>\keq$. No such dichotomy exists in $\emag(k)$, which shows a self-similar range of scales on $\ks < k < \ku^b$ with $\emag(k) \sim k^{-9/5}$, which is inconsistent with theories for Alfv\'enic turbulence \citep{Goldreich1995}, dynamic alignment \citep{Boldyrev2006}, and high-$\Rm$ tearing instabilities  \citep{Galishnikova2022_saturation_and_tearing,Dong2022_reconnection_mediated_cascade}. We report all empirical measurements of slopes in \ref{app:slopes}. Similar spectra have previously been measured in subsonic turbulence at high $\Re$ in \citep{Fielding2022_ISM_plasmoids}. Based on the physical size-scale of the plasmoids in the voids, we show the upper bound of $k_{*}$ on each of the panels. We see no significant spectral steepening on these scales. We further decompose the velocity into incompressible and compressible modes and show the spectra for those modes in Figure~\ref{fig:spectra}~(b). We find the compressible mode spectrum follows roughly $\ekin(k) \sim k^{-2}$ for all $k$, and the incompressible mode spectrum closely matches the total kinetic spectrum, with a slightly shallower spectrum $\ekin(k) \sim k^{-7/4}$ on the $k > \keq$ scales. This shows that the compressible modes are not passively tracing the incompressible modes, as is the standard result from previous compressible theories \citep{Lithwick2001_compressibleMHD}, but are potentially either transported non-locally via shocks across all scales, or undergo their own cascade. This is the first MHD simulation to resolve two kinetic energy cascades within a single simulation, i.e., $\ekin(k) \sim k^{-2}$ and $\ekin(k)\sim k^{-3/2}$, as well as a magnetic cascade, $\emag(k)\sim k^{-9/5}$, unexplained by any current MHD turbulence theory. The spectra we have shown are globally isotropic, which are measured in the absence of an imposed magnetic field. However, this does not mean that there is no local anisotropy, as expected for MHD turbulence. We provide an analysis of the local anisotropy in \ref{app:anisotropy}.

    \begin{figure}[h!]
        \centering
        \includegraphics[width=\linewidth]{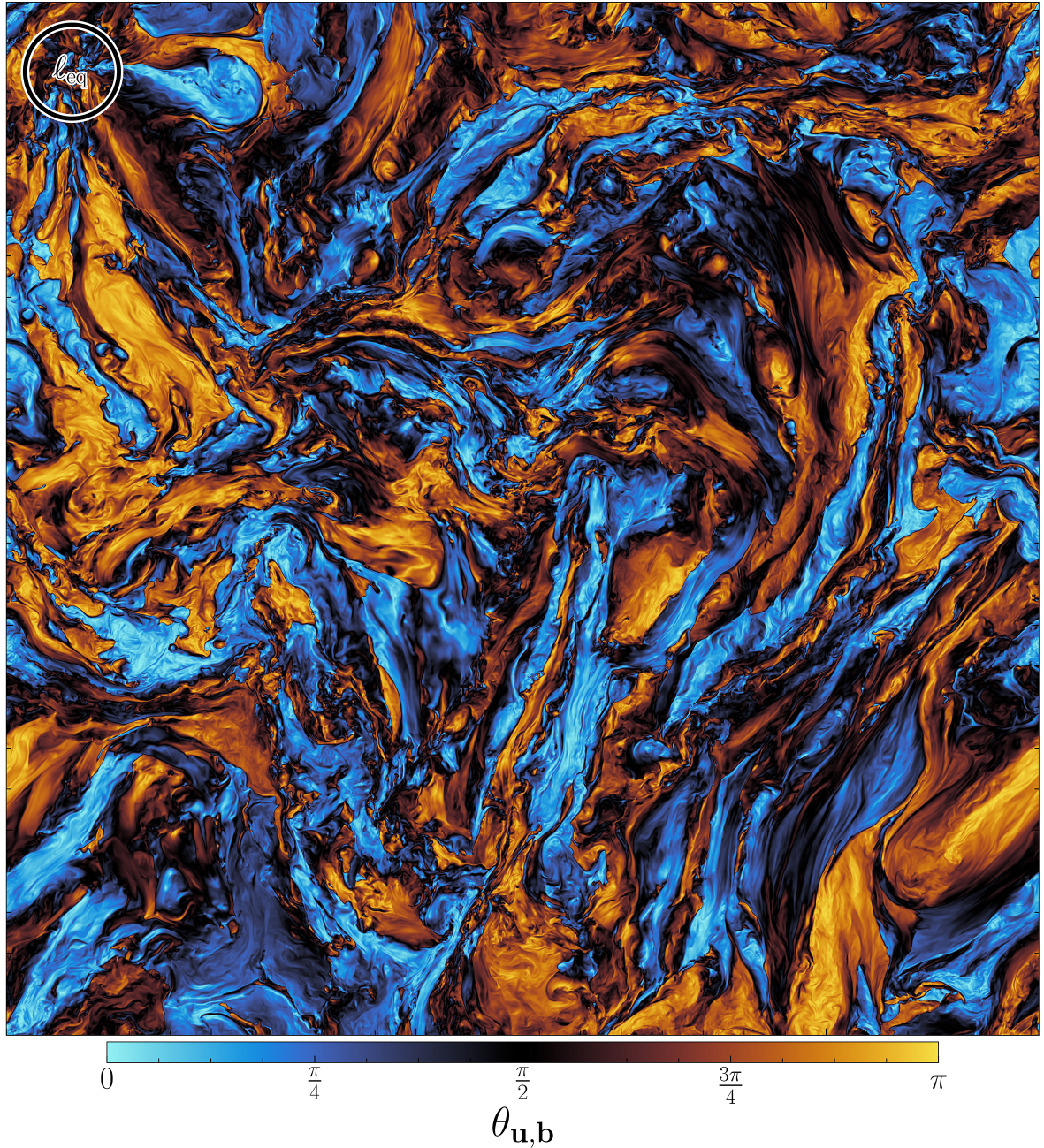}
        \caption{\textbf{The spatial structure of the alignment angle between $\bfu$ and $\bfb$.} A two-dimensional slice of the $\theta_{\bfu,\bfb}$ field, where parallel $\bfu$ and $\bfb$ are shown in blue  $\theta_{\bfu,\bfb}=0$, perpendicular  $\theta_{\bfu,\bfb}=\pi/2$ in black and anti-parallel $\theta_{\bfu,\bfb}=\pi$ in yellow. The size scale of the energy equipartition scale, $\ell_{\rm eq}$, is shown in the top left corner. The plasma is self-organized into volume-filling regions of parallel and anti-parallel $\bfu\propto \pm\bfb$ configurations. The perpendicular $\bfu$ and $\bfb$ configuration, where the nonlinearities in the turbulence are the strongest and hence the cascade is the fastest, is confined to the fractal interfaces between parallel and anti-parallel regions.}
        \label{fig:v_b_theta}
    \end{figure}

    \begin{figure}[h!]
        \centering        
        \includegraphics[width=\linewidth]{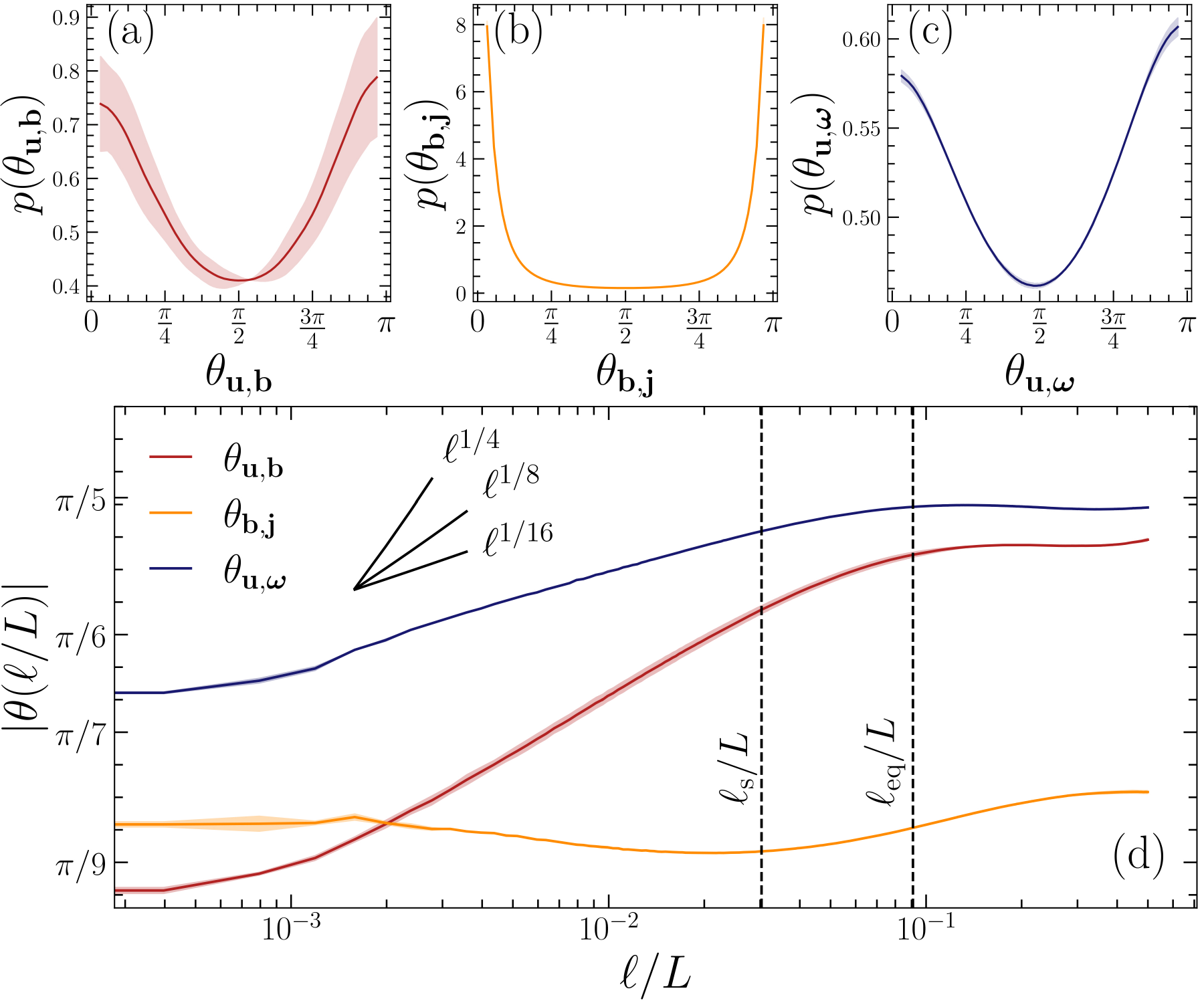}
        \caption{\textbf{The global and local alignment statistics, probing the self-organization into plasma relaxation states.} \textbf{(a)-(c)}: the one-point distribution functions of $\theta_{\bfu,\bfb}$, $\theta_{\bfb,\bfj}$ and $\theta_{\bfu,\bfo}$, all showing bimodal distribution functions peaked at the plasma relaxation states (either parallel or anti-parallel), $\bfu \propto \bfb \propto \bfj \propto \bfo$, inevitably reducing the strength of the nonlinearities that facilitate the turbulence \citep{Stribling1991_relaxation_processes,Banerjee2023_relaxed_states,Pecora2023_relaxation_in_magnetosheath}. \textbf{(d)} the scale-dependent absolute local angle between each of the field variables as a function of length scale. Both $|\theta_{\bfu,\bfb}| \sim \ell^{1/8}$ and $|\theta_{\bfu,\bfo}| \sim \ell^{1/16}$ show scale dependent alignment at smaller scales than the energy equipartition scale $\ell_{\rm eq}$, indicating that the nonlinearities in both the induction and momentum equation are becoming progressively weaker throughout the subsonic cascade. Consequently, this also limits the maintenance of the magnetic field to the large scales, as predicted for the turbulent dynamo saturation \citep{Schekochihin2002_saturation_evolution}. The relation, $|\theta_{\bfu,\bfb}| \sim \ell^{1/8}$ is inconsistent with the dynamical alignment prediction $|\theta_{\bfu,\bfb}| \sim \ell^{1/4}$ \citep{Boldyrev2006,Perez2009_dynamical_alignment_of_imbalanced_islands}, and is currently unexplained by any turbulence theory. Unlike the other alignments, $|\theta_{\bfb,\bfj}|$ exhibits scale-independent alignment, which means the turbulence is tending toward a globally force-free state ($\bfb \propto \pm\bfj$), suppressing the Lorentz force on all scales.}
        \label{fig:alignment_variables}
    \end{figure}

    \subsection{The alignment statistics of supersonic turbulence:} 
        An IK-type spectrum $\ekin(k) \sim k^{-3/2}$ has been associated with a depletion of the nonlinearities in the turbulence through the process of alignment \citep{Boldyrev2006,Mallet2015_refined_cb,Chernoglazov2021_alignment_SR_MHD,Banerjee2023_relaxed_states}, compared to the strongly nonlinear spectrum, $\sim k^{-5/3}$ \citep{Goldreich1995}. Alignment could be dynamical, from deformations of counter-propagating shear Alfv\'en wave packets \citep{Boldyrev2006,Perez2009_dynamical_alignment_of_imbalanced_islands}, or come from plasma relaxation, if the relaxation timescale is shorter than the nonlinear one. Solar wind and terrestrial magnetosheath studies suggest that this is indeed possible, but the relaxation process happens in localized regions rather than globally in the whole plasma \citep{Matthaeus2008_rapid_alignment,Pecora2023_relaxation_in_magnetosheath}. We therefore show a two-dimensional slice of the angle $\theta_{\bfu,\bfb}$ between $\bfu$ and $\bfb$, for the $10080^3$ simulation in Figure~\ref{fig:v_b_theta}, showing volume-filling $\bfu\propto\pm\bfb$ states, and volume-poor $\bfu\perp\bfb$ states, suggesting that local regions of the plasma indeed become very strongly aligned. To quantify this further, we show the global (panels a-c) and scale-dependent  alignment statistics (panel d) for $\bfu$, $\bfb$, $\bfj$ and $\bfo$ in Figure~\ref{fig:alignment_variables}, as well as field slices for the angles between $\bfu$ and $\bfb$, $\bfb$ and $\bfj$, and $\bfu$ and $\bfo$ in Figure~S5. In general, we find that $\bfu \propto \bfb \propto \bfj \propto \bfo$ states are preferred, with all $\theta$ probability density functions (a) - (c), showing strongly peaked bimodal distributions around parallel and anti-parallel configurations, hence we conclude that the turbulence is indeed undergoing a relaxation process. If $\bfu$ and $\bfb$ follow a process akin to dynamical alignment, then $|\theta_{\bfu,\bfb}(\ell)| \sim \ell^{1/4}$ \citep{Boldyrev2006}. We provide details about the exact definitions for this statistic in \ref{app:scale_dependent_dfn}. We provide the $|\theta(\ell)|$ functions for each of the relaxation variables, showing significant scale-dependent alignment in $|\theta_{\bfu,\bfb}(\ell)| \sim \ell^{1/8}$ and $|\theta_{\bfu,\bfo}(\ell)| \sim \ell^{1/16}$ compared to $|\theta_{\bfb,\bfj}(\ell)|$, which remains mostly scale independent and highly aligned across all of the scales in the turbulence, suggesting that the plasma is tending towards a global Taylor force-free state where $\bfb \propto \pm \bfj$. The outcome of $\bfu$ and $\bfb$ becoming progressively more aligned on $k > \keq$, is that the magnetic induction turns off on small scales, undoubtedly related to the saturation of the turbulent dynamo. The $|\theta_{\bfu,\bfb}(\ell)| \sim \ell^{1/8}$ scaling is inconsistent with the dynamic alignment prediction of incompressible MHD turbulence theory \citep{Boldyrev2006,Perez2009_dynamical_alignment_of_imbalanced_islands}, and hence a new theory is required for describing these scaling laws. 

    \begin{figure}
        \centering
        \includegraphics[width=\linewidth]{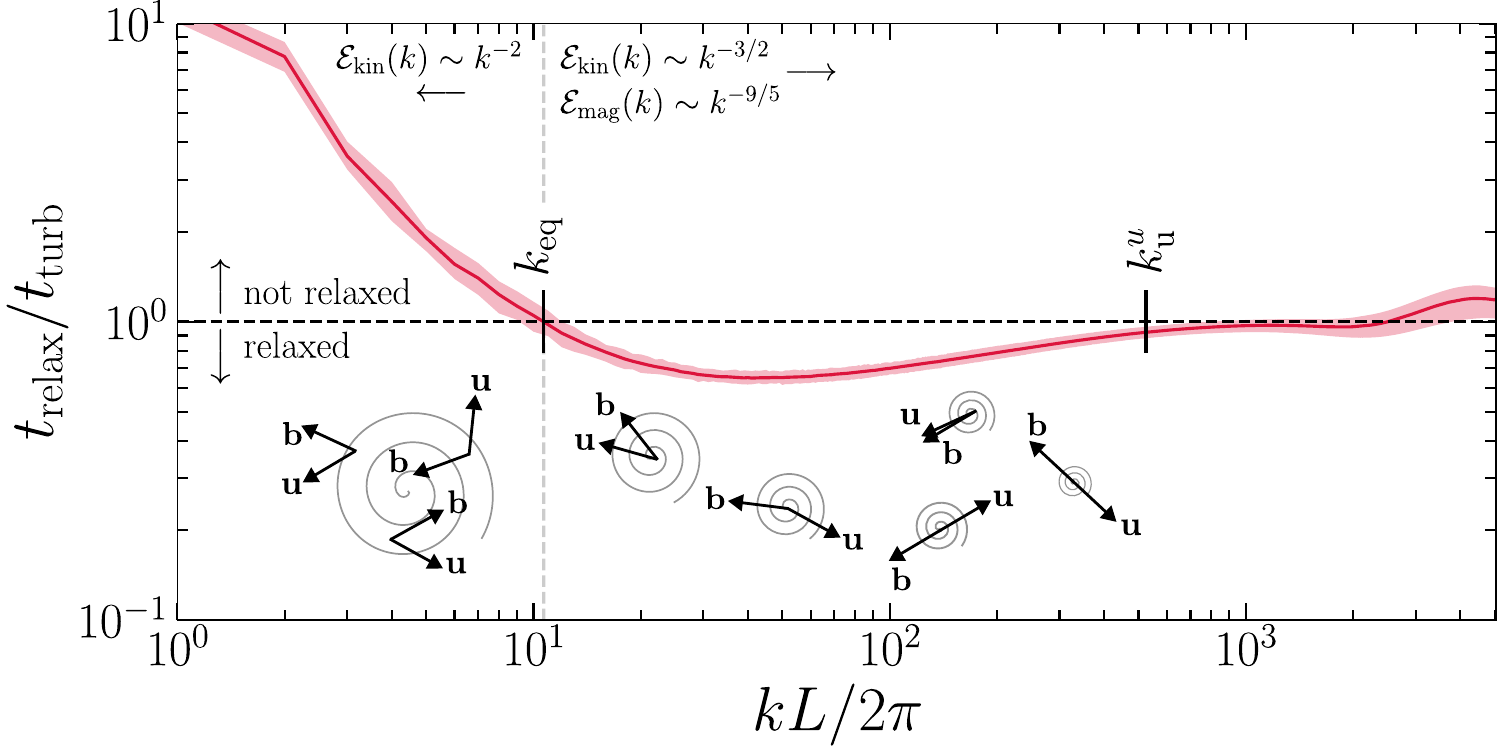}
        \caption{\textbf{The local competition between turbulence and plasma relaxation.} The ratio between the timescale for the turbulent velocity fluctuations $t_{\rm turb}$ and the approximate relaxation time $t_{\rm relax}$, which we assume is related to the dynamical timescale of the magnetic field. For modes larger than $\keq$, which correspond to the scales in the $\ekin(k) \sim k^{-3/2}$ and $\emag(k) \sim k^{-9/5}$ cascades, the relaxation is faster than the turbulent velocity timescale, $t_{\rm turb} > t_{\rm relax}$, hence the plasma is able to relax, tending to a $\bfu \propto \pm \bfb$ state and depleting the nonlinearities faster than the turbulent fluctuations are able to perturb it out of this state. We illustrate this with a cartoon drawing of the aligned $\bfu$ and $\bfb$ states, alongside swirling turbulent eddies. For modes smaller than $\keq$, the turbulent velocity timescale is shorter than the relaxation timescale, $t_{\rm turb} < t_{\rm relax}$, hence the turbulence randomizes the alignment angles and the nonlinear transfers dominate the spectrum, giving rise to the Burgers spectrum, $\ekin(k) \sim k^{-2}$. With the cartoon, we illustrate random orientations of the $\bfu$ and $\bfb$ states on these scales.}
        \label{fig:relaxation_time_scale}
    \end{figure}
        
        To understand what scales are able to relax in the plasma we now perform a scale-dependent timescale analysis. Because we know that we must treat the magnetic and velocity spectrum separately, as per the previous section, we must also treat the magnetic and velocity timescales separately, as opposed to combining them together into a single nonlinear timescale, as is the standard practice in driven turbulence studies \citep{Schekochihin2020_bias_review}. We approximate the scale-dependent relaxation timescale as $t_{\rm relax}(k) \sim t_{\rm A}(k) = 1 / [k \delta v_{\rm A}(k)]$, assuming that the process is related to the dynamical timescale of the magnetic field fluctuations, as has been found previously in  \citet{Dobrowolny1980_solar_wind_properties}. Because the turbulence is sourced through the velocity fluctuations, we compare this with the scale-dependent velocity fluctuation timescale $t_{\rm turb}(k) = 1/[k \delta u(k)]$, assuming that it is the source of the turbulent motions in the velocity that perturb each scale away from relaxation. This process can be qualitatively understood through the principle of vanishing nonlinear transfers \citep{Banerjee2023_relaxed_states}, where on each scale there is a competition between relaxation and the source of the turbulence, which, in a steady-state provides a deviation away from a relaxed state. We plot $t_{\rm relax}/t_{\rm turb}$ as a function of $k$ in Figure~\ref{fig:relaxation_time_scale}. For $k > \keq$, where $\ekin(k) \propto k^{-3/2}$ and $\delta u(k) / c_s < 1$, we find that $t_{\rm relax}/t_{\rm turb} < 1$, so the plasma is able to relax and deplete the nonlinearities faster than the turbulence is able to perturb the magnetic field out of a relaxed state. Conversely, for $k < \keq$ and $\delta u(k) /c_s > 1$, $t_{\rm relax}/t_{\rm turb} > 1$, and the turbulence is able to maintain strong nonlinear transfers, without significant weakening from relaxation. We therefore associate the shallow $\ekin(k)\sim k^{-3/2}$ spectrum with the depletion of nonlinearities via the local competition between plasma relaxation and the turbulence.

\subsection{The local anisotropy of MHD turbulence with no net magnetic flux}\label{app:anisotropy}
\begin{figure}
        \centering
        \includegraphics[width=\linewidth]{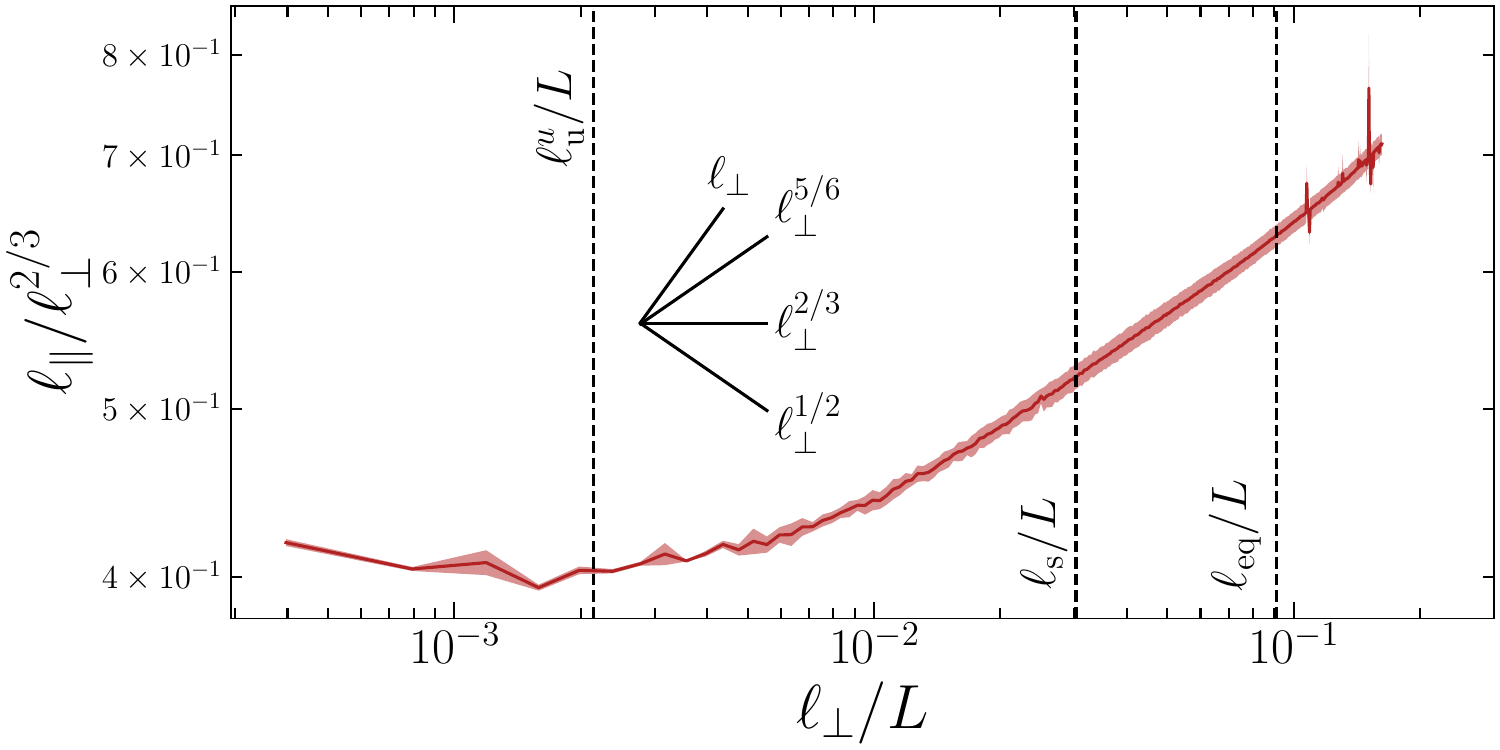}
        \caption{\textbf{Scale-dependent anisotropy in supersonic turbulence.} The scale-dependent anisotropy of the velocity from the $10080^3$ simulation, where $\ell_{\parallel}$ is the scale parallel to the local mean magnetic field and $\ell_{\perp}$ is the scale perpendicular. We compensate by the prediction for two-dimensional anisotropy for Alfv\'enic turbulence, $\ell_{\parallel} \sim \ell_{\perp}^{2/3}$ \citep{Goldreich1995}. We show the end of the subsonic cascade $\ell_{\rm u}^u$, the sonic scale $\ell_{\rm s}$, and the energy equipartition scale $\ell_{\rm eq}$. We also show the prediction for the three-dimensional anisotropy predictions from dynamical alignment, $\ell_{\parallel} \sim \ell_{\perp}^{1/2}$ \citep{Boldyrev2006,Mallet2017_anisotropy}. We find that neither the two-dimensional nor three-dimensional anisotropy relations aptly describe the supersonic turbulence, and the plasma tends towards $\ell_{\parallel} \sim \ell_{\perp}^{5/6}$ over a broad range of scales.}
        \label{fig:scale_dependent_anisotropy}
    \end{figure}

    Our simulations have no net magnetic flux, commonly referred to in the turbulent dynamo community as isotropic MHD. However, even though the magnetic and velocity fields are globally isotropic, with no preferential direction or orientation, the statistics in the coordinate frame of the local magnetic field need not be so, and a number of authors have shown that this is indeed the case, even for isotropic MHD \citep{Beresnyak2009_alignment,StOnge2020_weakly_collisional_dynamo,Fielding2022_ISM_plasmoids}. Following these studies, we compute two-dimensional $(\ell_{\parallel},\ell_{\perp})$ anisotropic structure functions for the velocity in the frame of the local mean magnetic field. This allows us to directly compare against the anisotropy predictions for incompressible MHD turbulence \citep{Goldreich1995,Boldyrev2006,Mallet2017_anisotropy}. As in \ref{app:scale_dependent_dfn}, we define $\widehat{\bfb_{\bfell}}$, and then the conditional, anisotropic structure functions,
    \begin{align}
        \delta\bfu^2(\ell_{\parallel} | \theta_{\bfell,\widehat{\bfb_{\bfell}}}) &= \Exp{[\bfu(\bfr) - \bfu(\bfr + \bfell)]^2 \Big|\;\; 0 \leq \theta_{\bfell,\widehat{\bfb_{\bfell}}} < \pi/18}{\bfell}, \\
        \delta\bfu^2(\ell_{\perp} | \theta_{\bfell,\widehat{\bfb_{\bfell}}}) &= \Exp{[\bfu(\bfr) - \bfu(\bfr + \bfell)]^2 \Big|\;\; 8\pi/18 < \theta_{\bfell,\widehat{\bfb_{\bfell}}} \leq \pi/2}{\bfell},
    \end{align}
    where 
    \begin{align}
         0 \leq \theta_{\bfell,\widehat{\bfb_{\bfell}}} = \arccos\frac{|\bfell\cdot\widehat{\bfb_{\bfell}}|}{|\bfell|} \leq \pi/2.
    \end{align}
    Next, we find $\ell_{\parallel}$ as a function of $\ell_{\perp}$ by matching the amplitudes between the two structure functions, e.g.
    \begin{align}
        \delta\bfu^2(\ell_{\parallel} | \theta_{\bfell,\widehat{\bfb_{\bfell}}}) &=
        \delta\bfu^2(\ell_{\perp} | \theta_{\bfell,\widehat{\bfb_{\bfell}}}),
    \end{align}
    and then matching the $\ell_{\parallel}$ for the corresponding $\ell_{\perp}$ that satisfies this relation \citep{StOnge2020_weakly_collisional_dynamo,Fielding2022_ISM_plasmoids}. We plot the $\ell_{\parallel}$ as a function of $\ell_{\perp}$ for the $10080^3$ simulation in Figure~\ref{fig:scale_dependent_anisotropy}, compensating $\ell_{\parallel}$ by $\ell_{\perp}^{2/3}$ to directly test if the anisotropy is consistent with the two-dimensional anisotropic predictions $\ell_{\parallel} \sim \ell_{\perp}^{2/3}$ \citep{Goldreich1995}. We find that indeed there is significant scale-dependent anisotropy, with $\ell_{\parallel} \sim \ell_{\perp}^{5/6}$, stronger scale-dependent anisotropy than the prediction for both two-dimensional anisotropy \citep{Goldreich1995} and three-dimensional anisotropy $\ell_{\parallel} \sim \ell_{\perp}^{1/2}$\citep{Boldyrev2006,Mallet2017_anisotropy}. This relation persists over a large range of scales, including both $\ell_{\rm eq}$ and $\ell_{\rm s}$. Deviations away from these relations have been found before in similar, (but incompressible), isotropic MHD dynamo experiments, at much lower plasma Reynolds numbers \citep{StOnge2020_weakly_collisional_dynamo}.

    \begin{figure}
        \centering
        \includegraphics[width=\linewidth]{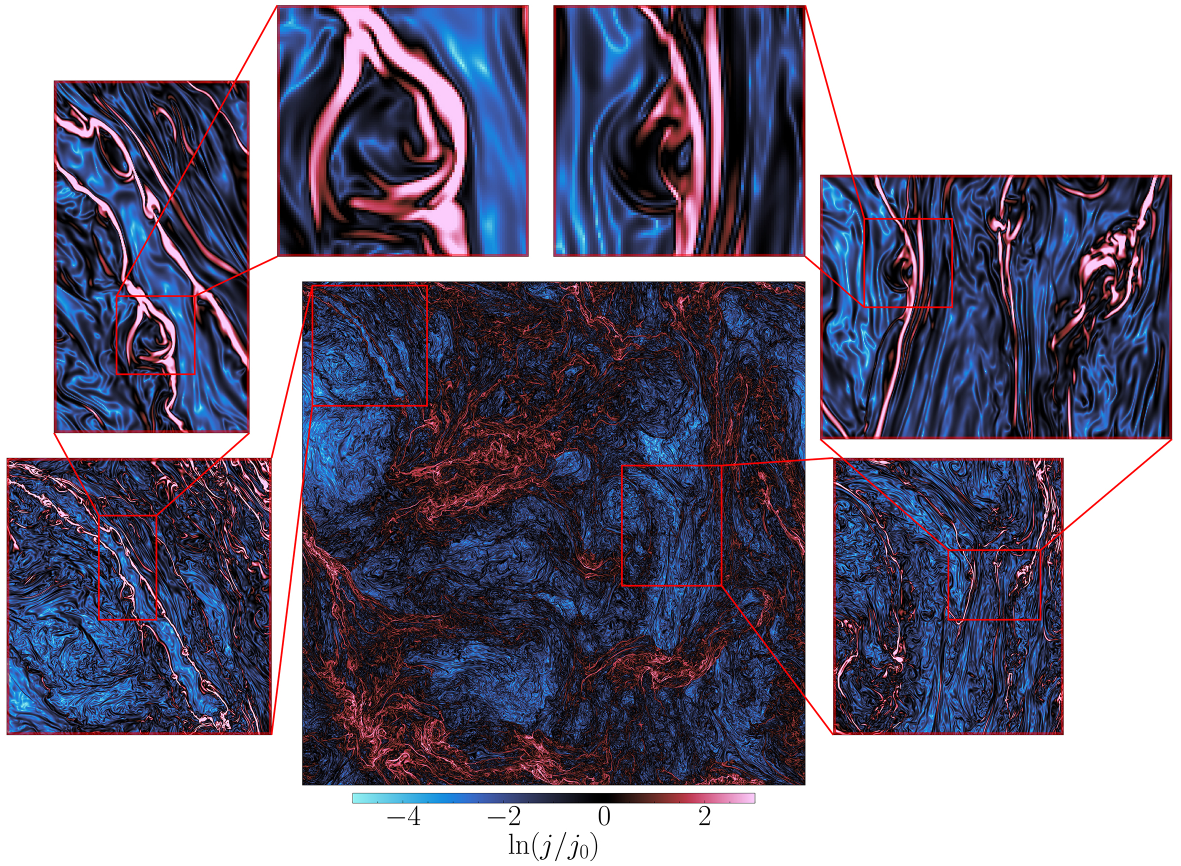}
        \caption{\textbf{Tearing instabilities forming in current sheets spread through the voids of supersonic MHD turbulence.} A two-dimensional logarithmic current density slice using the same color scheme as shown in the top of Figure~\ref{fig:slice_poster}. Three levels of zoom-ins reveals chains of plasmoid or flux tube structures appearing in two thin current sheets situated in the gas density voids. We zoom directly into one of the plasmoids appearing in each chain, with the maximum zoom-in revealing the entire cross-section of the plasmoid. The size scale of this region is $k=100$. These well-structured plasmoid chains appear in the most isolated sheets, away from the large, intense body of current that corresponds to the formation of a dense gas filament (see Figure~\ref{fig:slice_poster} for $\rho$ and $\bfj$ correlations), suggesting that disruptions events from the supersonic shocks moving through the turbulence prevent coherent formation of plasmoids everywhere in the supersonic plasma.}
        \label{fig:plasmoids}
    \end{figure}

\subsection{Tearing instabilities and plasmoid chains in current sheets}\label{app:plasmoids}
    Tearing instabilities are ubiquitous in reconnecting current sheets when the magnetic resistivity is small compared to the out-flow velocity from the reconnecting sheet \citep{Bhattacharjee2009_fast_reconnection,Uzdensky2010_fast_reconnection,Loureiro_2016_plasmoid_instability,Boldyrev2017_MHD_mediated_by_reconnection,Mallet2017_plasmoid_disruptions,Comisso2018_MHD_turbulence_plasmoid_regime,Boldyrev2020_tearing_mode_instability}. One can parameterize this ratio using the Lundquist number $S = \ell_{\rm sheet} v_{\rm A} / \eta$, where $\ell_{\rm sheet}$ is the length of the sheet, $v_{\rm A}$ is the characteristic plasma out-flow velocity of the sheet, and $\eta$ is the magnetic resistivity. For low $S$, the reconnection rate is given by $\epsilon_{\rm rec} \sim S^{-1/2}$, the slow reconnection regime, critically relying on the resistivity through $S$. However, for $S \gtrsim 10^5$, the current sheets may become unstable to tearing modes instabilities if the nonlinear timescale of the turbulence is longer than the growth timescale of the tearing modes \citep{Loureiro2017_reconnection_in_turbulence,Mallet2017_plasmoid_disruptions,Boldyrev2017_reconnection_in_turbulence,Comisso2018_MHD_turbulence_plasmoid_regime,Loureiro2020_reconnection_in_turbulence,Dong2022_reconnection_mediated_cascade}. This gives rise to regions of intense filamentary currents that spontaneously disrupt the sheet. This instability allows $\epsilon_{\rm rec}$ to become independent of $S$ \citep{Bhattacharjee2009_fast_reconnection,Uzdensky2010_fast_reconnection}, entering a fast reconnection regime, and exciting turbulent modes with a $\ekin(k) \sim \emag(k) \sim k^{-11/5}$ spectrum \citep{Dong2022_reconnection_mediated_cascade}. Based on Figure~\ref{fig:slice_poster}, $\ell_{\rm sheet} \approx \ell_0$, and using the $\emag(k)/\ekin(k)$ spectra in Figure~\ref{fig:spectra}, $v_{\rm A} \approx 0.1 \delta u$ on $\ell_0$. Hence, in our $10080^3$ simulations, where $\Rm \sim 3 \times 10^6$, $S \gtrsim 10^5$ on the current sheet length scale, as required for the fast reconnection regime.
    
    We search for tearing-type instabilities in our $10080^3$ simulations to understand what scales they may influence, in an effort to understand if they change the nature of the energy cascade in supersonic magnetohydrodynamic turbulence. We show a zoom-in of an isolated set of intense current sheets in Figure~\ref{fig:plasmoids}, going from the full $k L/2\pi=1$ mode domain down to the $k L/2\pi=100$ mode in top-most panels, revealing chains of plasmoid-like structures, which are the generic outcome of the current sheet becoming tearing unstable \citep{Bhattacharjee2009_fast_reconnection,Uzdensky2010_fast_reconnection,Loureiro_2016_plasmoid_instability,Boldyrev2017_MHD_mediated_by_reconnection,Mallet2017_plasmoid_disruptions,Comisso2018_MHD_turbulence_plasmoid_regime,Boldyrev2020_tearing_mode_instability}, developing in the sheets. Both sheets are found within the gas density voids in the turbulence, but several other disrupted, fractal sheets can be found in the shocked gas (see Figure~\ref{fig:slice_poster} to find that shocked gas corresponds to the large current clumps in Figure~\ref{fig:plasmoids}). The plasmoids are small, no larger than $k L/2\pi \sim 100$, placing them on the $k_*L/2\pi \gtrsim 100$ modes in the energy spectra shown in Figure~\ref{fig:spectra}, which are resolved in our simulations. Even though these plasmoids are present, it is not clear that the reconnection-mediated energy cascade \citep{Dong2022_reconnection_mediated_cascade} translates perfectly to the supersonic regime, where the current sheet structure itself may be disrupted by large-scale shocks in both the velocity and gas density. In force balance, which the shocked gas tends toward, $c_s^2\bm{\nabla}\rho = \bfj \times \bfb/\mu_0$ for our isothermal plasma \citep{Robertson2018,Mocz2018}, which means that the current sheets form on orthogonal flux surfaces to $\bm{\nabla}\rho$, $\bfj\cdot\bm{\nabla}\rho = 0$. Hence, the structure of the current sheets are sensitive to the volume-filling properties of $\bm{\nabla}\rho$, which makes the role of plasmoids in the supersonic regime significantly different from the subsonic, incompressible regime \citep{Dong2022_reconnection_mediated_cascade}.

\section{Summary and conclusions}\label{sec:discussion}
        By running the world's largest driven supersonic MHD turbulence simulations, we have revealed the existence of two scale-separated $\ekin(k)$ cascades -- a Burgers-type cascade ($\sim k^{-2}$), which hosts supersonic, kinetic-energy-dominated motions, and an IK-type cascade ($\sim k^{-3/2}$), which hosts subsonic, magnetically-dominated motions. Many authors have measured a $k^{-3/2}$ spectrum before \citep{Maron2001,Muller2005_k32_total_spectrum,Perez2009_dynamical_alignment_of_imbalanced_islands,Bowen2018_steep_solar_wind_spectra,Dong2022_reconnection_mediated_cascade}, which is attributed to strong, three-dimensional-anisotropic MHD turbulence undergoing dynamical alignment. However, we show explicitly in Figure~\ref{fig:alignment_variables} that our scale-dependent alignment $|\theta_{\bfu,\bfb}|\sim\ell^{1/8}$ is inconsistent with the dynamical alignment prediction $|\theta_{\bfu,\bfb}| \sim \ell^{1/4}$ \citep{Boldyrev2006,Perez2009_dynamical_alignment_of_imbalanced_islands}. Moreover, based on the preference for $\bfu \propto \bfb \propto \bfj \propto \bfo$, we argue that on the modes where $\delta u / c_s < 1$ and $k > \keq$, it is plasma relaxation in local patches of the turbulence that gives rise to the weaker $\sim k^{-3/2}$ spectrum, as has been conjectured previously, motivated by measurements of $\sim 1\;\text{AU}$ regions of the solar wind and the terrestrial magnetosheath \citep{Matthaeus2008_rapid_alignment,Osman2011_solar_wind_alignment,Pecora2023_relaxation_in_magnetosheath}. This means that both the magnetic and kinetic energy cascades happen inhomogeneously through the plasma, and fastest along the boundary surfaces between aligned and anti-aligned $\bfu$ and $\bfb$ regions. Understanding the size scale and stability of these relaxed regions \citep[discussed previously in the appendix of ][]{Hosking2020_tangled_field_stats} will be important for determining the role of turbulence in all astrophysical systems, since a growing stable, relaxed region can halt the cascade completely. 
    
        As for $\emag(k)$, we find a single $\sim k^{-9/5}$ cascade that is significantly different from the $\ekin(k)$ cascades, and thus necessitates separate theoretical treatment \citep{Grete2021_as_a_matter_of_tension}, even on the scales where $\ekin(k)\sim\emag(k)$, as shown in Figure~\ref{fig:spectra}~(d). At a fundamental level, this means we cannot combine the velocity and magnetic fluctuations into a single nonlinear timescale. Indeed, different cascade exponents between the energy spectra have been found in the solar wind before, with the magnetic spectrum becoming as steep as $\emag(k) \sim k^{-2}$, which has been associated with intermittent structures in the plasma \citep{Bowen2018_steep_solar_wind_spectra,Dunn2023_steep_solar_wind_spectra}. We want to further highlight that this spectrum is emergent on the scales that are relaxing faster than being perturbed by the turbulence (see Figure~\ref{fig:relaxation_time_scale}), and hence we cannot ignore that $\sim k^{-9/5}$ is likely a feature of the $\bfb \propto \pm\bfj$ and $\bfu \propto \pm\bfb$ relaxation states. With the two cascades in the kinetic energy, the single cascade in magnetic energy, the new scale-dependent alignment relations, and relaxation state results, this study describes a new paradigm for understanding supersonic MHD turbulence, which we hope stimulates many further fundamental, theoretical investigations.

        Finally, the results in this study should be brought into the context of the saturation of the $\Pm \sim 1$ fluctuation dynamo, which is, after all, what we are simulating in this high $\Re$ and $\Rm$ turbulent regime. Moreover, because our simulation resolves both super and subsonic $k$ modes, and the dynamics in both of the regimes are scale separated, this simulation is very useful for understanding the saturation of the subsonic turbulent dynamo, where we can confidently say that any effect from the turbulent driving will be negligible due to the large scale separation between $k_0$ and $\ks$. Panel~(d) in Figure~\ref{fig:spectra}, shows that all of the subsonic $k$ have a stationary state where $\emag(k) \gtrsim \ekin(k)$, indicating that if there is enough scale separation between the driving and subsonic cascade, the subsonic modes should all reach a strongly magnetized state. This state is important for the dynamo saturation mechanism, because as the $k$ modes become magnetized, $t_{\rm A} < t_{\rm turb}$, which means that $\bfb$ can evolve faster than $\bfu$, and dynamically configure into a relaxed, aligned state, as we show in Figure~\ref{fig:relaxation_time_scale}. This paints a picture for the saturation of the dynamo based upon local relaxation, that we will now describe, but with the intention to only describe the broad picture and leave the details for future works. Consider the kinematic dynamo, where $\emag \ll \ekin$. From \citet{Kazantsev1968} theory, or even simple scaling arguments, the growth rate is associated with dynamical timescale on the viscous scale, $t_{\nu} \sim  \ell_{\nu}^2/\nu \sim \ell_{\nu}/u_{\nu}$ \citep{Kazantsev1968,Schekochihin2004_dynamo,Galishnikova2022_saturation_and_tearing,Kriel2022_kinematic_dynamo_scales,Beattie2023_bulk_viscosity}. Once $\emag(\ell_{\nu}) \sim \ekin(\ell_{\nu})$, $t_{\rm A} \sim t_{\nu}$ on $\ell_{\nu}$, and there is a transition from the kinematic to the nonlinear dynamo regime. This implies $\delta u_{\nu} \sim \delta b_{\nu}$ (where $\delta b_{\nu}$ is the magnetic field fluctuation on $\ell_{\nu}$), and means that $\delta b_{\nu}$ is able to relax to a state $\delta u_{\nu} \parallel \pm \delta b_{\nu}$. This turns off the magnetic flux generation on this scale, since $\nabla\times(\bfu \times \bfb)$ can be turned off geometrically if $\delta u_{\nu} \parallel \pm \delta b_{\nu}$. Now the next eddy on $\ell > \ell_{\nu}$ is responsible for the magnetic flux generation, but as each successively larger scale grows and $\delta u_{\ell} \sim \delta b_{\ell}$, then that scale is able to relax into a $\delta u \parallel \pm \delta b$ state, moving the magnetic flux generation to larger and large scales, until it is being sourced on the largest scales $\sim \ell_{\rm eq}$, where $\delta u$ and $\delta b$ are aligned the least. This is a dynamical process, happening on $t_{\rm A}$ timescales, making the dynamo saturation process dynamical and not resistive. It requires no additional effective turbulent resistivity suppressing the growth \citep{Schober2015_saturation_of_turbulent_dynamo,Xu2016_dynamo}, which is inconsistent with direct measurements of the magnetic energy equation showing that resistivity terms decrease during saturation when compared to the kinematic stage \citep{Seta2021_supersonic_saturation}. The saturated state is therefore one that is aligned on small scales, with magnetic flux being generated on large scales, resulting in a classically structured turbulent magnetic spectrum, peaked at low $k$, as shown in Figure~\ref{fig:spectra}, rather than a Kazantsev one.

\section{Methods}\label{sec:methods}
\subsection{Numerical simulations}\label{app:simulations}
\paragraph{Basic numerical code:} 
    We use a heavily modified version of the magnetohydrodynamical (MHD) code \textsc{flash} \citep{Fryxell2000,Dubey2008}. Our code uses a highly-optimized, hybrid-precision \citep{Federrath2021}, positivity-preserving, second-order MUSCL-Hancock HLL5R Riemann scheme \citep{Bouchut2010,Waagan2011} to solve the compressible, ideal, MHD fluid equations in three dimensions,
    \begin{align}
    \partial_t \rho + \nabla\cdot\left(\rho \bfu\right) = 0,& \label{eq:continuity}\\
    \partial_t\!\left(\rho \bfu\right) + \nabla\cdot\left(\rho\bfu\!\otimes\!\bfu + p\mathbb{I} - \frac{1}{\mu_0}\bfb\!\otimes\!\bfb \right) = \rho \bm{ f},&\label{eq:momentum} \\
    \partial_t \bfb  + \nabla\cdot(\bfu\otimes\bfb - \bfb\otimes\bfu) = 0,& \label{eq:induction}\\ 
    \nabla\cdot\bfb = 0,& \label{eq:divb} \\
    p = \cs^2\rho+ \frac{1}{2\mu_0}\bfb \cdot \bfb,& \label{eq:pressure}
    \end{align}
    where $\rho$, $\bfu$, $\bfb$ and $\mu_0$ are the gas density, the velocity and magnetic fields, and the magnetic permittivity, respectively. Equation~\ref{eq:pressure} relates the scalar pressure $p$ to $\rho$ via the isothermal equation of state with constant sound speed $c_s$, as well as the pressure contribution from the magnetic field. We work in units $c_s = \rho_0 = \mu_0 = L = 1$, where $\rho_0$ is the mean gas density and $L$ is the characteristic length scale of the system, such that $L^3 = \mathcal{V} = 1$ is the volume. We discretize the equations over a triply periodic domain of $[-L/2, L/2]$ in each dimension, with grid resolutions $2520^3$, $5040^3$ and $10080^3$ -- the largest grids in the world for simulations of this fluid turbulence regime. In order to drive turbulence, a turbulent forcing term $\bm{f}$ is applied in the momentum equation (details below). This set of equations including the forcing term is the standard approach in modeling driven, magnetized turbulence. These calculations were only possible as part of a large-scale high performance computing project, large-scale project 10391, at Leibniz Supercomputing Centre in Garching, Germany. They were run on the supercomputer SuperMUC-NG. For the $10080^3$ simulation, and the power-spectrum calculations, we utilized close to 140,000 compute cores, and close to 80M-core hours. 

\paragraph{Turbulent driving:} 
    We choose to drive the turbulence with a turbulent Mach number of $\delta u /c_s \approx 4$ to ensure that we resolve a sufficient range of both supersonic $\delta u > c_s$ and subsonic $\delta u < c_s$ scales \citep{Federrath2021}. We apply a non-helical stochastic forcing term $\bm{f}$ in Equation~\ref{eq:momentum}, following an Ornstein-Uhlenbeck stochastic process \citep{Eswaran1988_forcing_numerical_scheme,Schmidt2009,Federrath2010_solendoidal_versus_compressive}, using the \textsc{TurbGen} turbulent forcing module \citep{Federrath2010_solendoidal_versus_compressive,Federrath2022_turbulence_driving_module}. The forcing is constructed in Fourier space such that kinetic energy is injected at the smallest wavenumbers, peaking at $\ell_0^{-1} = k_0L/2\pi = 2$ and tending to zero parabolically in the interval $1\leq kL/2\pi\leq3$, allowing for self-consistent development of turbulence on smaller scales, $kL/2\pi>3$, as routinely performed in turbulence box studies. To replenish the large-scale compressible modes and shocks, we decompose $\bm{f}$ into its incompressible ($\nabla\cdot\bm{f}=0$) and compressible ($\nabla\times\bm{f}=0$) modes \citep{Federrath2010_solendoidal_versus_compressive}, and drive the turbulence with equal amounts of energy in each of the modes. 

\paragraph{Initial conditions and hierarchical interpolation:}\label{app:init_conditions}
    We initialize $\rho(x,y,z) = \rho_0$ and $\bfu = \bm{0}$. The total magnetic field $\bfb = \bfb_0 + \bfb_{\rm turb}$ is composed of both a mean (external or guide) field $\bfb_0$ and turbulent $\bfb_{\rm turb}$ component. The $\bfb_{\rm turb}$ evolves self-consistently with the MHD turbulence via Equation~\ref{eq:induction}, whereas $\bfb_0$ remains constant throughout the simulation due to magnetic flux conservation. For our simulations, $\bfb_0 = 0$, and only the isotropic, turbulent magnetic field remains, $\bfb = \bfb_{\rm turb}$. Regardless of the initial field amplitude, the same small-scale dynamo saturation is reached \citep{Beattie2023_growth_or_decay}. The same also holds for different seed magnetic fields \citep{Seta2020_seed_magnetic_field}. Hence, given enough integration time, we can initialize a magnetic field with any initial structure and amplitude and be confident that it will result in the same saturation. Such that we do not use computational resources on the fast or nonlinear dynamo stages \citep{Federrath2016_dynamo,Rincon2019_dynamo_theories}, we initialize the magnetic field amplitude and structure close to the saturated state. From our previous experiments at lower resolutions, this is $\emag/\ekin \approx 1/4$ (or $\Ma \approx 2$, where $\Ma = \delta u / \delta v_{\rm A}$ is the Alfv\'en Mach number), and with a significant amount of power at all $k$. We find that a uniform initial $\emag(k)$, with sufficiently high $k$ modes included relaxes very quickly to the saturated dynamo state. Hence we use a simple, isotropic uniform top-hat spectrum within $1 \leq kL/2\pi \leq 50$, which generates as a divergence-free, static, random field using \textsc{TurbGen} \citep{Federrath2010_solendoidal_versus_compressive,Federrath2022_turbulence_driving_module}, where the total integral of the spectrum is set such that the magnetic energy is $\emag \approx \ekin/4$, and $\ekin = \rho_0 \delta u^2 / 2$ is estimated directly from the driving.  

    Driven MHD turbulence in this regime takes $\approx (1-2)t_0$, where $t_0 = \ell/\delta u$ is the turbulent turnover time on the outer scale, to shed the influence of its initial conditions and establish a stationary state \citep{Federrath2010_solendoidal_versus_compressive,Price2010_grid_versus_SPH,Beattie2022_spdf}. To avoid expending compute resources on simulating this transient state, we only apply the previously discussed initial conditions to the $2520^3$ simulation. For the remaining $5040^3$ and $10080^3$ simulations, we interpolate the initial conditions hierarchically from the simulations with lower resolutions, i.e., we initialize the $5040^3$ simulation with linearly interpolated initial conditions from the $t \approx 2t_0$ state of the $2520^3$ simulation and the $t \approx 3t_0$ state of the $5040^3$ simulation for the $10080^3$ simulation. We use linear interpolation to preserve $\nabla\cdot\bfb = 0$ between grid interpolations. It takes a tiny fraction of $t_0$, $t \sim \Re^{-1/2}t_0$, to populate the new modes after the interpolation onto the higher-resolution grid. Hence this provides an adequate method for minimizing the amount of compute time spent making the $5040^3$ and $10080^3$ simulations stationary.
    
\paragraph{Estimating the Reynolds numbers:} 
    Our numerical model is an implicit large eddy simulation (ILES), which relies upon the spatial discretisation to supply the numerical viscosity and resistivity as a fluid closure model. Recently, a detailed characterization of our code's numerical viscous and resistive properties, specifically for turbulent boxes, has been performed by comparing the ILES model with direct numerical simulations (DNS), which have explicit viscous and resistive operators in Equation~\ref{eq:momentum} and Equation~\ref{eq:induction}, respectively \citep{Kriel2022_kinematic_dynamo_scales,Grete2023_as_a_matter_of_dynamical_range,Shivakumar2023_numerical_dissipation}. \citet{Shivakumar2023_numerical_dissipation} derived empirical models for transforming grid resolution $N_{\rm grid}$ into $\Re$ and $\Rm$. For supersonic MHD turbulence, they find, $\Re = (N_{\rm grid}/N_{\Re})^{p_{\Re}}$, where $p_{\Re} \in [1.5, 2.0]$ and $N_{\Re} \in [0.8, 4.4]$ and $\Rm = (N_{\rm grid}/N_{\Rm})^{p_{\Rm}}$, where $p_{\Rm} \in [1.2, 1.6]$ and $N_{\Rm} \in [0.1, 0.7]$. For our three $N_{\rm grid}$, $2520$, $5040$ and $10080$, this gives $\Re_{2 \rm k} \in [1.8 \times 10^5, 3.3 \times 10^5]$, $\Re_{5 \rm k} \in [5\times 10^5 , 1.3 \times 10^6]$, $\Re_{10k} \in [1.4\times 10^6, 5.3\times 10^6]$ and $\Rm_{2\rm k} \in [1.9\times10^5, 4.9\times10^5]$, $\Rm_{5 \rm k} \in [4.4\times10^5, 1.5\times10^6]$, $\Rm_{10\rm k} \in [1\times10^6, 4.5\times10^6]$, where the $2\rm k$, $5\rm k$ and $10\rm k$ subscripts correspond to the $N_{\rm grid} = 2520$, $5040$ and $10080$ simulations, respectively. When we report the Reynolds numbers in the main text, we report the average between the bounds placed on each of the dimensionless plasma numbers.

\paragraph{Data structure and domain decomposition:} 
    \textsc{flash} uses a block-structured parallelization. Each 3D computational block is distributed onto one single compute core. For the $10080^3$ simulation, it uses $168\!\times\!210\!\times\!210$ cells per block, resulting in $(168\!\times\!60, 210\!\times\!48, 210\!\times\!48) = (10080, 10080, 10080)$ cells in each spatial direction, for a total of $60\!\times\!48\!\times\!48=138,\!240$~cores used in the $10080^3$ run. The $5040^3$ and $2520^3$ simulations, which are used to check numerical convergence of statistical quantities, have a block structure $(84\!\times\!60, 210\!\times\!24, 210\!\times\!24) = (5040, 5040, 5040)$, using $34,\!560$~cores, and $(42\!\times\!60, 210\!\times\!12, 210\!\times\!12) = (2520 , 2520, 2520)$, using $8,\!640$~cores, respectively.

\paragraph{File I/O:} 
    \textsc{flash} is parallelised with \textsc{mpi}. File I/O is based on the \textsc{hdf}\oldstylenums{5} library. Since our \textbf{} runs will use $138,\!240$~cores (3024~compute nodes on SuperMUC-NG) and produce approximately $100$ output files with $29\,\mathrm{TB}$ each (approximately $3.0\,\mathrm{PB}$ in total), efficient file I/O is extremely important. In order to achieve the highest efficiency when reading and writing these huge files, we use parallel-\textsc{hdf}\oldstylenums{5} together with a split-file approach. In this approach each core writes simultaneously to disk, grouping data from 288~cores together into a total of 504 files per output dump. This proved to be an extremely efficient method, providing an I/O throughput that is close to the physical maximum of approximately $200\,\mathrm{GB/s}$ reachable on the SuperMUC-NG \textsc{/scratch} file system. The net effect is that I/O only takes about 3--4 minutes to read or write a $28\,\mathrm{TB}$ checkpoint file, such that it consumes only a minor fraction of the resources compared to the integration of the MHD fluid equations.

\paragraph{Major code optimizations:} 
    Our version of \textsc{flash} is highly optimized for solving large-scale hydrodynamical and MHD problems \citep{Federrath2021}. Specifically, the number of stored 3D fields are reduced to the bare minimum required for these simulations (only the gas density and three velocity and magnetic field components are stored). All calls to the equation of state routines are performed inline, directly in the Riemann solver. The code is precision hybridized such that all fluid variables are stored in single precision (4~bytes per floating-point number), but critical operations are performed in double-precision arithmetic (8~bytes per floating-point number), which retains the accuracy of the full double-precision computations. These efforts significantly reduce the computational time and the required amount of \textsc{mpi} communication, as well as the overall memory consumption. In addition, the single-precision operations benefit from a higher SIMD count and lower cache occupancy, for a further parallel speedup. As a result the code is almost $4\times$ faster and requires $4.1\times$ less memory than the \textsc{flash} public release, while retaining the full accuracy. Previous studies have further characterized the performance and comparison of our code with the public \textsc{flash} version \citep{Cielo2020_code_scaling}.

\subsection{Definition of energy spectra and turbulent scales}\label{app:turbulent_scales}
    \paragraph{Energy spectra:}
    The magnetic energy spectrum is defined as 
    \begin{align}
        \emag(k) = \frac{1}{2\mu_0}\int\d{\Omega_k}\;  \Tilde{\bfb}(\mathbf{k})\Tilde{\bfb}^{\dagger}(\mathbf{k}) 4\pi k^2,
    \end{align}
    and kinetic energy spectrum is defined as,
    \begin{align}
        \ekin(k) = \frac{\rho_0}{2}\int \d{\Omega_k}\; \Tilde{\bfu}(\mathbf{k}) \Tilde{\bfu}^{\dagger}(\mathbf{k}) 4\pi k^2,
    \end{align}
    where the tilde indicates the Fourier transform of the underlying field variable, dagger the complex conjugate and the $\int \d{\Omega_k}4\pi k^2$ is the shell integral over fixed $k$ shells, producing isotropic, one-dimensional energy spectra. Note that other definitions of the kinetic energy spectrum have been used in the literature, which define $\mathbf{w} = \sqrt{\rho}\bfu$ and then take the square Fourier transform of $\mathbf{w}$ to construct the kinetic energy spectrum \citep{Federrath2010_solendoidal_versus_compressive,Kritsuk2007,Federrath2013_universality,Grete2017_shell_models_for_CMHD,Grete2020_as_a_matter_of_state,Grete2023_as_a_matter_of_dynamical_range}. However, we pick the simplest definition of $\ekin(k)$ to allow us to more easily compare with theories of incompressible turbulence \citep{Iroshnikov_1965_IK_turb,Kraichnan1965_IKturb,Goldreich1995,Boldyrev2006}, and even compressible theories \citep{Bhattacharjee_1998_weakly_compressible_solar_wind,Lithwick2001_compressibleMHD,Federrath2021}, which adopt the same $\ekin(k)$ definition as we do in this study.

    \paragraph{Inner and outer scales:}
         We use the following definitions for the inner and outer scales of the turbulent cascades, using $u$ and $b$ superscripts to differentiate between the kinetic energy and magnetic energy scales, respectively. For both the $\emag(k)$ and $\ekin(k)$, the outer scale is directly related to the integral or correlation scale of the energy spectrum, 
        \begin{align}
            \frac{\kcor L}{2\pi} = \frac{\displaystyle \int\d{k}\,\mathcal{E}(k)}{\displaystyle \int\d{k}\, (kL/2\pi)^{-1}\mathcal{E}(k)},  
        \end{align}
        which for $\ekin(k)$ closely tracks the driving scale $\kcor^u \sim k_0$, but for $\emag(k)$ depends on a range of parameters, like the growth stage of the dynamo and the strength of the large-scale magnetic field \citep{Beattie2022energy_balance,Beattie2023_growth_or_decay}. For our $10080^3$ simulation, $\kcor$ of the magnetic field is $\kcor^b L/2\pi \sim 10$, highlighting how the magnetic field is intrinsically a small-scale field. For the inner scale we take the maximum of the $(kL/2\pi)^2\mathcal{E}(k)$ spectrum,
        \begin{align}
            \frac{\ku L}{2\pi} = \left(\frac{\max\left\{(kL/2\pi)^2\mathcal{E}(k)\d{k}\right\}}{\displaystyle \int\d{k}\,\mathcal{E}(k)}\right)^{1/2},
        \end{align}
        which probes the smallest scale of the magnetic and velocity gradients, since, e.g.\ for the velocity, $k^2\ekin(k)\d{k} \sim k^2 u^2 \sim (\nabla\otimes\bfu)^2$, defining the end of the turbulent cascade and the start of diffusion-dominated scales. We show both of these scales in panel (a) and (c) in Figure~\ref{fig:spectra}.
        
    \begin{figure}
        \centering
        \includegraphics[width=\linewidth]{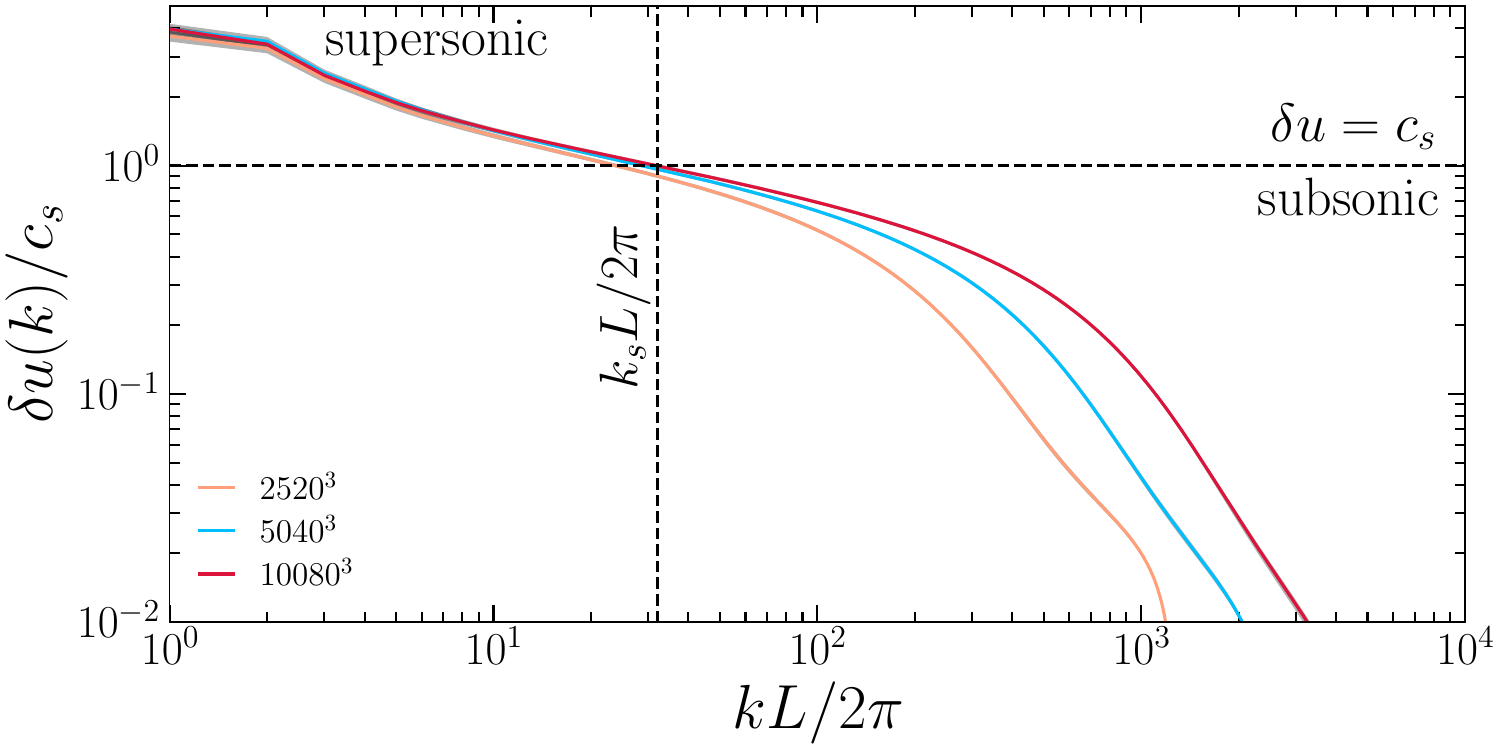}
        \caption{\textbf{Velocity dispersion as a function of wavenumber and the sonic scale.} The root mean square (rms) velocity normalized by the sound speed, $\delta u /c_s$, as a function of wavenumber $k$. The wavenumber where $\delta u = c_s$, the sonic scale $\ks L/2\pi$, is indicated with a horizontal, black, dashed line, with $\delta u > c_s$, corresponding to supersonic rms velocities, and $\delta u < c_s$ to subsonic rms velocities. The sonic scale, $\ks L/2\pi = 32 \pm 2$ for the $10080^{3}$ simulation, is indicated with a vertical dashed line.}
        \label{fig:k_space_vel_dispersion}
    \end{figure}

    \paragraph{The sonic scale:} 
        The first transition that we investigate is the $k$-space sonic scale, denoted as $\ks L/2\pi$. Using Parseval's theorem, we calculate the rms velocity as a function of scale through the relation
        \begin{align}\label{eq:k_space_vel}
        \frac{\delta u(k)}{c_s} = \left(\frac{2}{\rho_0 }\int^{\infty}_{k}\d{k'}\,\ekin(k')\right)^{1/2},
        \end{align}
        where $\ks L/2\pi$ is the $k$ mode where $\delta u(k_{\rm s}) = c_s$ \citep{Federrath2010_solendoidal_versus_compressive,Federrath2012,Federrath2021}. We present this calculation in Figure~\ref{fig:k_space_vel_dispersion} and determine the $\ks L / 2\pi$ root for the sonic transition
        \begin{align}\label{eq:sonic_scale}
            \delta u(k)/c_s - 1 = 0,
        \end{align}
        which is indicated by the dashed line. For the $10080^3$ simulation, we find $\ks L/2\pi = 32 \pm 2$, for $5040^3$, $\ks L/2\pi = 29 \pm 2$ and $2520^3$, $\ks L/2\pi = 24 \pm 2$. For $k>\ks$, the plasma becomes subsonic with $\delta u(k)/c_s < 1$, and for $k<\ks$, the plasma is supersonic with $\delta u(k)/c_s > 1$. This is marked by the horizontal black dashed line. As previously found, there is a smooth transition between these two flow regimes rather than a sharp discontinuity \citep{Federrath2021}. 
    
    \paragraph{Energy equipartition scale:}
        The second transition that we study is the transition from kinetic energy dominated turbulence, $\ekin(k) > \emag(k)$ to magnetic energy dominated turbulence, $\emag(k) > \ekin(k)$, which is equivalent to comparing the turbulent Alfv\'en timescale, $t_{\rm A} = \ell/\delta v_{\rm A}(\ell)$, where $\delta v_{\rm A}$ is the rms Alfv\'en velocity of the plasma, with the turbulent velocity timescale $t_{\rm turb} = \ell/\delta u(\ell)$. If $\ekin(k) > \emag(k)$, then $t_{\rm turb} < t_{\rm A}$ and vice versa for $\ekin(k) < \emag(k)$. For strong guide field turbulence, this has been previously called the MHD scale \citep{Goldreich1995,Lithwick2001_compressibleMHD}, but we use the more general energy equipartition scale nomenclature, $\keq L/2\pi$, as in the main text. Even though comparing these timescales looks like a calculation about critical balance, since both timescales are describing intrinsically nonlinear fluctuations, this is not a probe for weak versus strong turbulence \citep{Perez2008_weak_and_strong_turb,Meyrand2016_strong_to_weak_transition}. To determine $\keq L /2\pi$ we find the root of $\emag(k)/\ekin(k) - 1 = 0$. We plot the full $\emag(k)/\ekin(k)$ spectrum and $\keq L /2\pi$ mode in panel (d) of Figure~\ref{fig:spectra}. For the $10080^3$ simulation, we find $\keq L/2\pi = 10.6 \pm 0.7$, for $5040^3$, $\keq L/2\pi = 11.9 \pm 0.7$ and $2520^3$, $\keq L/2\pi = 13.1 \pm 0.6$. 

\subsection{Definitions of scale-dependent alignment structure functions}\label{app:scale_dependent_dfn}
    To compute the scale-dependent alignment structure function shown in Figure~\ref{fig:alignment_variables} (d), we first define our increments,
    \begin{align}
        \bm{\delta}\bfu &= \bfu(\bfr) - \bfu(\bfr + \bfell), \\
        \bm{\delta}\bfb &= \bfb(\bfr) - \bfb(\bfr+\bfell),
    \end{align}  
    for separation vector $\bm{\ell}$. Next, we define a local mean magnetic field direction,
    \begin{align}
        \widehat{\bfb}_{\bfell} = \frac{\bfb(\bfr) + \bfb(\bfr + \bfell)}{\|\bfb(\bfr) + \bfb(\bfr + \bfell)\|},
    \end{align}
    and then find the perpendicular component to the local field for each of the fluid variables, e.g. for $\bfu$ and $\bfb$,
    \begin{align}
        \bm{\delta}\bfu_{\perp} &= \bm{\delta}\bfu - (\bm{\delta}\bfu\cdot \widehat{\bfb}_{\bfell})\widehat{\bfb}_{\bfell}, \\
        \bm{\delta}\bfb_{\perp} &= \bm{\delta}\bfb - (\bm{\delta}\bfb\cdot\widehat{\bfb}_{\bfell})\widehat{\bfb}_{\bfell},
    \end{align}
    which is the standard definition for these quantities \citep{Chernoglazov2021_alignment_SR_MHD,Dong2022_reconnection_mediated_cascade}. Next we construct the ratio between first-order structure functions,
    \begin{align}
    |\theta_{\bfu,\bfb}(\ell)| \sim |\sin\theta_{\bfu,\bfb}(\ell)| = \frac{\Exp{|\bm{\delta}\bfu_{\perp} \times \bm{\delta}\bfb_{\perp}|}{\ell}}{\Exp{ | \bm{\delta}\bfu_{\perp} | | \bm{\delta}\bfb_{\perp} |}{\ell}}.
    \end{align}
    We do this for each relaxation variable pair, $\bfu$ and $\bfb$, $\bfj$ and $\bfb$, and lastly $\bfo$ and $\bfu$. We use $2 \times 10^{12}$ sampling pairs to ensure that the structure functions are converged at all scales \citep{Federrath2021}. Furthermore, we construct the structure functions across a number of realizations in the stationary state, and then time-average the structure function to produce Figure~\ref{fig:alignment_variables}. 

\backmatter

\bmhead{Supplementary information}

\section{Empirical measurements for the slopes of the energy spectra}\label{app:slopes}
    In Figure~\ref{fig:spectra} and the corresponding section we provide tilde slopes, accompanied by compensations in each of the panels. Here we directly report the slopes utilizing weighted linear least squares on the linearized counterpart of the model $\mathcal{E}(k) = \beta_0 k^{\beta_1}$. For the weights, we use the 1$\sigma$ from the time-averaged spectra. For the kinetic energy spectra we partition the $k$ domain into supersonic scales within the supersonic cascade, $4 \leq k_{\rm super}L/2\pi \leq \keq L/2\pi$, and subsonic scales, within the subsonic cascade $\ks L/2\pi = 33 \leq k_{\rm sub}L/2\pi \leq 10 \ks L/2\pi = 330$. We find, in general, our choices for the fit domain do not have a large impact on the exact values, as long as we pick scales within the cascades. For $\ekin(k_{\rm super})$ we find $\beta_1 = -2.01 \pm 0.03$, and for $\ekin(k_{\rm sub})$, $\beta_1 = -1.465 \pm 0.002$, close to the tilde values we present in the main text, $\sim -2$ and $\sim -3/2$, respectively. Performing the same analysis for the compressible $\ekin^{\rm comp}$ and solenoidal $\ekin^{\rm sol}$ mode kinetic energy spectra we find for $\ekin^{\rm comp}(k_{\rm super})$ $\beta_1 = -2.05 \pm 0.04$, for $\ekin^{\rm comp}(k_{\rm sub})$, $\beta_1 = -1.971 \pm 0.001$, $\ekin^{\rm sol}(k_{\rm super})$ $\beta_1 = -1.97 \pm 0.05$ and $\ekin^{\rm sol}(k_{\rm sub})$ $\beta_1 = -1.425 \pm 0.001$, reinforcing that the compressible modes follow a single spectrum $\sim k^{-2}$, not passively tracing the incompressible modes, and the incompressible modes capture the supersonic-to-subsonic dichotomy. We do the same fits to the magnetic spectra over the single domain $80 \leq  k L/2\pi \leq 250$, and find $\beta_1 = -1.798 \pm 0.001$, consistent with the $\sim k^{9/5}$ scaling we compensate the spectra by in the main text.

    \begin{figure}
            \centering
            \includegraphics[width=\linewidth]{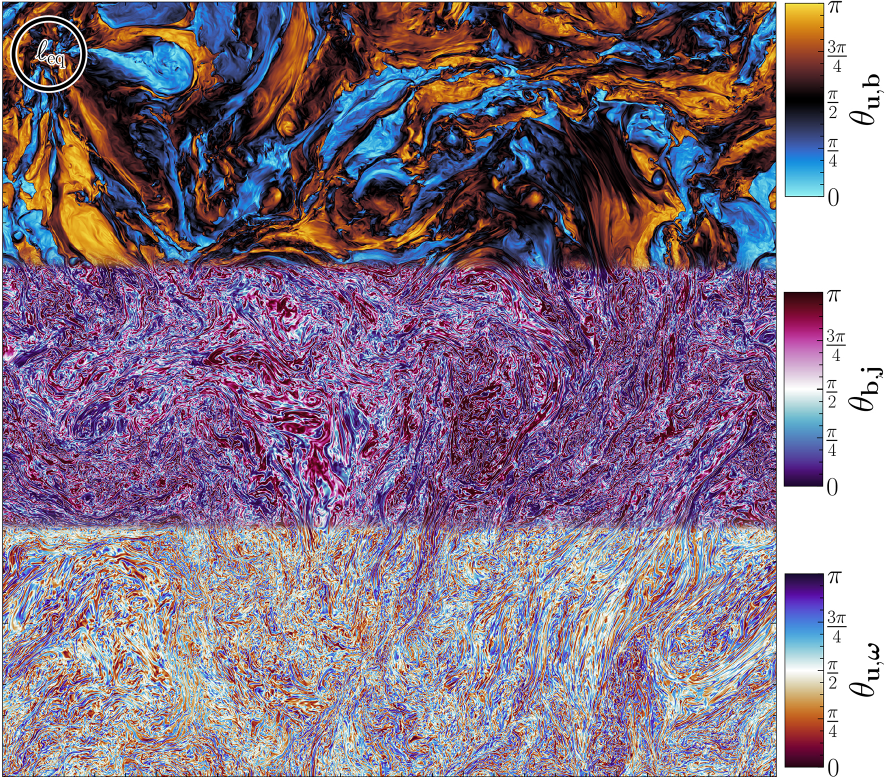}
            \caption{\textbf{Visualizations of aligned variables showing prominent locally relaxed regions on different scales.} Two-dimensional slices of the $\theta_{\bfu,\bfb}$ (top), $\theta_{\bfb,\bfj}$ (middle) and $\theta_{\bfu,\bfo}$ (bottom) fields in the $10080^3$ simulation. The $\theta_{\bfu,\bfb}$ field shows large-scale fluctuations of $\bfu\propto\pm\bfb$, roughly equal to the annotated energy equipartition scale, $\ell_{\rm eq}$ , which is the scale where the scale-dependent alignment starts, as shown in the structure function in panel (d) of Figure~\ref{fig:alignment_variables}. The other $\bfb$ and $\bfj$, and $\bfu$ and $\bfo$ alignments are on much smaller scales, and show weaker scale-dependence compared to $\bfu$ and $\bfb$.}
            \label{fig:alignment_fields}
    \end{figure}

\subsection{Basics principles of classical, global plasma relaxation}\label{app:relaxation}
    A relaxed plasma is one that is in a minimum energy state for a given set of constraints. For an ideal MHD plasma, the constraints, which are the ideal quadratic invariants of the plasma, are the cross helicity, $H_c = \Exp{\bfu\cdot\bfb}{\V}$, magnetic helicity $H_m = \Exp{\bfa\cdot\bfb}{\V}$ and the total energy $\mathcal{E} = \ekin + \emag$, where $\bfb = \nabla\times\bfa$ is the magnetic vector potential and $\V$ the volume. Finding the minimum energy state is then akin to using the variational principle to find the minima of the energy functional, 
    \begin{align}
        \delta \left(\frac{1}{\V}\iiint\d{\V}\;\; \left[ \mathcal{E} - 2\lambda_1 (\bfu\cdot\bfb) - 2\lambda_2 (\bfa\cdot\bfb) \right] \right)= 0,
    \end{align}
    where $\lambda_1$ and $\lambda_2$ are the Lagrangian multipliers and $\delta$ is the variational operator. The Lagrange multipliers are completely determined by the initial values of the constraints. Constructing $\mathcal{E}$ directly from Equation~\ref{eq:momentum}-\ref{eq:induction}, one can solve the above equation to show that the minimum energy state is,
    \begin{align}
        \bfu = \lambda_2\bfb, \quad\quad \bfj = \frac{(1-\lambda_2)^2}{\lambda_1}\bfb, \quad\quad \bfu = \frac{(1-\lambda_2)^2}{\lambda_1}\bfo,
    \end{align}
    \citep{Pecora2023_relaxation_in_magnetosheath}. Depending on the sign of $\lambda_1$ and $\lambda_2$, the minimum energy state for fixed $H_c$ and $H_m$ is the alignment or anti-alignment of $\bfu$, $\bfb$, $\bfj$ and $\bfo$. Note that this is a global minimization, over the whole volume $\V$, which we show here as the \textit{modus operandi} for problems of this nature. Alignment of these vectors matter \citep{Banerjee2023_relaxed_states}. Alignment between $\bfu$ and $\bfo$ reduces the nonlinearity that gives rise to the hydrodynamical turbulence, because $\bfu\cdot\nabla\otimes\bfu = \nabla u^2/2 + \bfu\times\bfo$. For incompressible, strong Alfv\'enic turbulence models \citep{Boldyrev2006,Perez2009_dynamical_alignment_of_imbalanced_islands}, the nonlinear timescale of the turbulence becomes slower in the presence of $\bfu$ and $\bfb$ alignment, and the nonlinearity parameter, weaker \citep{Mallet2015_refined_cb,Chernoglazov2021_alignment_SR_MHD}. The alignment between $\bfj$ and $\bfb$ suppresses the Lorentz force $\bfj \times \bfb$, creating force-free, Taylor relaxed states. For the turbulent dynamo, the magnetic energy flux is sourced from the induction term in Equation~\ref{eq:induction}, $\nabla\times(\bfu\times\bfb) = \nabla\cdot(\bfu\otimes\bfb - \bfb\otimes\bfu)$, and hence both the generation and maintenance of a magnetic field is sensitive to the scale-dependent alignment of $\bfu$ and $\bfb$. 
    
    As we show in Figure~\ref{fig:alignment_variables}, the one-point probability density functions of the angles between $\bfu$, $\bfb$, $\bfj$ and $\bfo$ variables show a strong preference for aligned, relaxed states. In Figure~S5 we show a two-dimensional slice of the underlying angle fields, revealing the structure that gives rise to one-point probability density functions. Similar to the magnetosheath discussed in \citep{Pecora2023_relaxation_in_magnetosheath}, we find that the turbulence ``cellularises" into locally relaxed patches, unlike what we have described above, which is a global relaxed state. The outer scale of the patches changing for the different variables, and hence the relations derived above will not strictly hold for our plasma. The largest scale patches are in $\theta_{\bfu\cdot\bfb}$, which have fluctuations on scales of order $\ell_{\rm eq}/L$, which we directly annotate on the plot. This is completely consistent with what we have presented in the main text, where scales below $\ell_{\rm eq}/L$ are able to relax faster than be perturbed out of relaxation from the turbulence, shown in Figure~\ref{fig:relaxation_time_scale}, and that it is scales below $\ell_{\rm eq}/L$ where the $\ekin(k)\sim k^{-3/2}$ and $\emag(k) \sim k^{-9/5}$ spectra emerge, which we argue is due to relaxation. Both the other angle fields exhibit much smaller-scale features.

\bmhead{Acknowledgements}
    We acknowledge the useful discussions with Drummond Fielding, Alexander Chernoglazov and Andrey Beresnyak on the local anisotropy and alignment structure functions, and the more general discussions about this work with Philip Mocz, Neco Kriel, Bart Ripperda and Chris Thompson. \textbf{Funding:} J.~R.~B.~acknowledges financial support from the Australian National University, via the Deakin PhD and Dean's Higher Degree Research (theoretical physics) Scholarships and the Australian Government via the Australian Government Research Training Program Fee-Offset Scholarship and the Australian Capital Territory Government funded Fulbright scholarship. J.~R.~B., C.~F., R.~S.~K. and S.~C further acknowledge high-performance computing resources provided by the Leibniz Rechenzentrum and the Gauss Centre for Supercomputing grants~pr32lo, pr73fi and GCS large-scale project~10391.
    C.F.~acknowledges funding by the Australian Research Council (Discovery Projects grant~DP230102280), and the Australia-Germany Joint Research Cooperation Scheme (UA-DAAD). C.F.~further acknowledges high-performance computing resources provided by the Leibniz Rechenzentrum and the Gauss Centre for Supercomputing (grants~pr32lo and pr48pi), the Australian National Computational Infrastructure (grant~ek9) and the Pawsey Supercomputing Centre (project~pawsey0810) in the framework of the National Computational Merit Allocation Scheme and the ANU Merit Allocation Scheme.
    R.~S.~K.~acknowledges support from the European Research Council via the ERC Synergy Grant “ECOGAL” (project ID 855130), from the German Excellence Strategy via the Heidelberg Cluster of Excellence (EXC 2181 - 390900948) “STRUCTURES”, and from the German Ministry for Economic Affairs and Climate Action in project “MAINN” (funding ID 50OO2206). R.~S.~K.~also thanks for local computing resources provided by the Ministry of Science, Research and the Arts (MWK) of {\em The L\"{a}nd} through bwHPC and the German Science Foundation (DFG) through grant INST 35/1134-1 FUGG and 35/1597-1 FUGG, and also for data storage at SDS@hd funded through grants INST 35/1314-1 FUGG and INST 35/1503-1 FUGG. J.~.R.~B. and A.~B further acknowledge the support from NSF Award 2206756. 
    \textbf{Author Contributions:} J.~R.~B.~led the entirety of the project, including the GCS large-scale project~10391, running the simulations, co-developing the \textsc{FLASH} code and analysis programs used in this study, and led the writing and ideas presented in the manuscript. C.~F.~co-led the GCS large-scale project~10391, is the lead developer of the \textsc{FLASH} code and the analysis pipelines used in the study, and contributed to the ideas presented in this study and drafting of the manuscript. R.~S.~K.~co-led the GCS large-scale project~10391, and contributed to the ideas presented in this study and drafting of the manuscript. S.~C~provided invaluable technical advice and assistance during the GCS large-scale project proposal and during the run time of the simulations, provided support visualizing the large datasets, and contributed to the ideas presented in this study and drafting of the manuscript. A.~B.~contributed to the ideas presented in this study and drafting of the manuscript.
    \textbf{Competing interests:} We declare no competing interests.
    \textbf{License information:} Copyright ©2024 the authors, some rights reserved. 

\bibliographystyle{mnras.bst}
\bibliography{sn-bibliography} 

\end{document}

%% file: header.tex
\newcommand{\bfu}{\mathbf{u}}
\newcommand{\bfa}{\mathbf{a}}
\newcommand{\bfr}{\mathbf{r}}
\newcommand{\bfell}{\bm{\ell}}
\newcommand{\bfb}{\mathbf{b}}
\newcommand{\bfj}{\mathbf{j}}
\newcommand{\bfo}{\bm{\omega}}
\newcommand{\cs}{c_s}

\newcommand{\Ma}{\mathcal{M}_{\rm A}}
\newcommand{\ekin}{\mathcal{E}_{\rm kin}}
\newcommand{\emag}{\mathcal{E}_{\rm mag}}

\newcommand{\kperp}{k_{\perp}}
\newcommand{\Rm}{\mathrm{Rm}}
\renewcommand{\Re}{\mathrm{Re}}
\newcommand{\Pm}{\mathrm{Pm}}
\newcommand{\V}{\mathcal{V}}
\newcommand{\Exp}[2]{\left\langle{#1}\right\rangle_{#2}}
\renewcommand{\d}[1]{\ensuremath{\operatorname{d}\!{#1}}}

\newcommand{\keq}{k_{\rm eq}}
\newcommand{\kcor}{k_{\rm cor}}
\newcommand{\ks}{k_{\rm s}}
\newcommand{\ku}{k_{\rm u}}

%% file: sn-article.bbl
\begin{thebibliography}{}
\makeatletter
\relax
\def\mn@urlcharsother{\let\do\@makeother \do\$\do\&\do\#\do\^\do\_\do\%\do\~}
\def\mn@doi{\begingroup\mn@urlcharsother \@ifnextchar [ {\mn@doi@}
  {\mn@doi@[]}}
\def\mn@doi@[#1]#2{\def\@tempa{#1}\ifx\@tempa\@empty \href
  {http://dx.doi.org/#2} {doi:#2}\else \href {http://dx.doi.org/#2} {#1}\fi
  \endgroup}
\def\mn@eprint#1#2{\mn@eprint@#1:#2::\@nil}
\def\mn@eprint@arXiv#1{\href {http://arxiv.org/abs/#1} {{\tt arXiv:#1}}}
\def\mn@eprint@dblp#1{\href {http://dblp.uni-trier.de/rec/bibtex/#1.xml}
  {dblp:#1}}
\def\mn@eprint@#1:#2:#3:#4\@nil{\def\@tempa {#1}\def\@tempb {#2}\def\@tempc
  {#3}\ifx \@tempc \@empty \let \@tempc \@tempb \let \@tempb \@tempa \fi \ifx
  \@tempb \@empty \def\@tempb {arXiv}\fi \@ifundefined
  {mn@eprint@\@tempb}{\@tempb:\@tempc}{\expandafter \expandafter \csname
  mn@eprint@\@tempb\endcsname \expandafter{\@tempc}}}

\bibitem[\protect\citeauthoryear{{Banerjee}, {Halder}  \& {Pan}}{{Banerjee}
  et~al.}{2023}]{Banerjee2023_relaxed_states}
{Banerjee} S.,  {Halder} A.,   {Pan} N.,  2023, \mn@doi [Physical Review E]
  {10.1103/PhysRevE.107.L043201}, \href
  {https://ui.adsabs.harvard.edu/abs/2023PhRvE.107d3201B} {107, L043201}

\bibitem[\protect\citeauthoryear{Beattie, Federrath, Klessen  \&
  Schneider}{Beattie et~al.}{2019}]{Beattie2019b}
Beattie J.~R.,  Federrath C.,  Klessen R.~S.,   Schneider N.,  2019, \mn@doi
  [The Monthly Notices of The Royal Astronomical Society]
  {10.1093/mnras/stz1853}, 488, 2493

\bibitem[\protect\citeauthoryear{{Beattie}, {Krumholz}, {Skalidis},
  {Federrath}, {Seta}, {Crocker}, {Mocz}  \& {Kriel}}{{Beattie}
  et~al.}{2022a}]{Beattie2022energy_balance}
{Beattie} J.~R.,  {Krumholz} M.~R.,  {Skalidis} R.,  {Federrath} C.,  {Seta}
  A.,  {Crocker} R.~M.,  {Mocz} P.,   {Kriel} N.,  2022a, \mn@doi [The Monthly
  Notices of The Royal Astronomical Society] {10.1093/mnras/stac2099}, \href
  {https://ui.adsabs.harvard.edu/abs/2022MNRAS.515.5267B} {515, 5267}

\bibitem[\protect\citeauthoryear{{Beattie}, {Mocz}, {Federrath}  \&
  {Klessen}}{{Beattie} et~al.}{2022b}]{Beattie2022_spdf}
{Beattie} J.~R.,  {Mocz} P.,  {Federrath} C.,   {Klessen} R.~S.,  2022b,
  \mn@doi [The Monthly Notices of The Royal Astronomical Society]
  {10.1093/mnras/stac3005}, \href
  {https://ui.adsabs.harvard.edu/abs/2022MNRAS.517.5003B} {517, 5003}

\bibitem[\protect\citeauthoryear{{Beattie}, {Federrath}, {Kriel}, {Hew}  \&
  {Bhattacharjee}}{{Beattie} et~al.}{2023a}]{Beattie2023_bulk_viscosity}
{Beattie} J.~R.,  {Federrath} C.,  {Kriel} N.,  {Hew} J. K.~J.,
  {Bhattacharjee} A.,  2023a, \mn@doi [arXiv e-prints]
  {10.48550/arXiv.2312.03984}, \href
  {https://ui.adsabs.harvard.edu/abs/2023arXiv231203984B} {p. arXiv:2312.03984}

\bibitem[\protect\citeauthoryear{{Beattie}, {Federrath}, {Kriel}, {Mocz}  \&
  {Seta}}{{Beattie} et~al.}{2023b}]{Beattie2023_growth_or_decay}
{Beattie} J.~R.,  {Federrath} C.,  {Kriel} N.,  {Mocz} P.,   {Seta} A.,  2023b,
  \mn@doi [The Monthly Notices of The Royal Astronomical Society]
  {10.1093/mnras/stad1863}, \href
  {https://ui.adsabs.harvard.edu/abs/2023MNRAS.524.3201B} {524, 3201}

\bibitem[\protect\citeauthoryear{{Beresnyak}}{{Beresnyak}}{2014}]{Beresnyak2014_4k_incomp_sim}
{Beresnyak} A.,  2014, \mn@doi [The Astrophysical Journal Letters]
  {10.1088/2041-8205/784/2/L20}, \href
  {https://ui.adsabs.harvard.edu/abs/2014ApJ...784L..20B} {784, L20}

\bibitem[\protect\citeauthoryear{{Beresnyak} \& {Lazarian}}{{Beresnyak} \&
  {Lazarian}}{2009}]{Beresnyak2009_alignment}
{Beresnyak} A.,  {Lazarian} A.,  2009, \mn@doi [The Astrophysical Journal]
  {10.1088/0004-637X/702/2/1190}, \href
  {https://ui.adsabs.harvard.edu/abs/2009ApJ...702.1190B} {702, 1190}

\bibitem[\protect\citeauthoryear{{Bhattacharjee}, {Ng}  \&
  {Spangler}}{{Bhattacharjee}
  et~al.}{1998}]{Bhattacharjee_1998_weakly_compressible_solar_wind}
{Bhattacharjee} A.,  {Ng} C.~S.,   {Spangler} S.~R.,  1998, \mn@doi [The
  Astrophysical Journal] {10.1086/305184}, \href
  {https://ui.adsabs.harvard.edu/abs/1998ApJ...494..409B} {494, 409}

\bibitem[\protect\citeauthoryear{{Bhattacharjee}, {Huang}, {Yang}  \&
  {Rogers}}{{Bhattacharjee} et~al.}{2009}]{Bhattacharjee2009_fast_reconnection}
{Bhattacharjee} A.,  {Huang} Y.-M.,  {Yang} H.,   {Rogers} B.,  2009, \mn@doi
  [Physics of Plasmas] {10.1063/1.3264103}, \href
  {https://ui.adsabs.harvard.edu/abs/2009PhPl...16k2102B} {16, 112102}

\bibitem[\protect\citeauthoryear{Boldyrev}{Boldyrev}{2006}]{Boldyrev2006}
Boldyrev S.,  2006, \mn@doi [Physical Review Letters]
  {10.1103/PhysRevLett.96.115002}, 96, 115002

\bibitem[\protect\citeauthoryear{{Boldyrev} \& {Loureiro}}{{Boldyrev} \&
  {Loureiro}}{2017a}]{Boldyrev2017_MHD_mediated_by_reconnection}
{Boldyrev} S.,  {Loureiro} N.~F.,  2017a, \mn@doi [The Astrophysical Journal]
  {10.3847/1538-4357/aa7d02}, \href
  {https://ui.adsabs.harvard.edu/abs/2017ApJ...844..125B} {844, 125}

\bibitem[\protect\citeauthoryear{{Boldyrev} \& {Loureiro}}{{Boldyrev} \&
  {Loureiro}}{2017b}]{Boldyrev2017_reconnection_in_turbulence}
{Boldyrev} S.,  {Loureiro} N.~F.,  2017b, \mn@doi [The Astrophysical Journal]
  {10.3847/1538-4357/aa7d02}, \href
  {https://ui.adsabs.harvard.edu/abs/2017ApJ...844..125B} {844, 125}

\bibitem[\protect\citeauthoryear{Boldyrev \& Loureiro}{Boldyrev \&
  Loureiro}{2020}]{Boldyrev2020_tearing_mode_instability}
Boldyrev S.,  Loureiro N.~F.,  2020, \mn@doi [Journal of Geophysical Research:
  Space Physics] {https://doi.org/10.1029/2020JA028185}, 125, e2020JA028185

\bibitem[\protect\citeauthoryear{Bouchut, Klingenberg  \& Waagan}{Bouchut
  et~al.}{2010}]{Bouchut2010}
Bouchut F.,  Klingenberg C.,   Waagan K.,  2010, \mn@doi [Numerische
  Mathematik] {10.1007/s00211-010-0289-4}, 115, 647

\bibitem[\protect\citeauthoryear{{Bowen}, {Mallet}, {Bonnell}  \&
  {Bale}}{{Bowen} et~al.}{2018}]{Bowen2018_steep_solar_wind_spectra}
{Bowen} T.~A.,  {Mallet} A.,  {Bonnell} J.~W.,   {Bale} S.~D.,  2018, \mn@doi
  [The Astrophysical Journal] {10.3847/1538-4357/aad95b}, \href
  {https://ui.adsabs.harvard.edu/abs/2018ApJ...865...45B} {865, 45}

\bibitem[\protect\citeauthoryear{{Brandenburg} \& {Ntormousi}}{{Brandenburg} \&
  {Ntormousi}}{2023}]{Brandenburg2023_galactic_dynamo_review}
{Brandenburg} A.,  {Ntormousi} E.,  2023, \mn@doi [Annu. Rev. Astron.
  Astrophys.] {10.1146/annurev-astro-071221-052807}, \href
  {https://ui.adsabs.harvard.edu/abs/2023ARA&A..61..561B} {61, 561}

\bibitem[\protect\citeauthoryear{{Br{\"u}ggen} \& {Vazza}}{{Br{\"u}ggen} \&
  {Vazza}}{2015}]{Bruggen2015_ICM_turbulence}
{Br{\"u}ggen} M.,  {Vazza} F.,  2015, in {Lazarian} A.,  {de Gouveia Dal Pino}
  E.~M.,   {Melioli} C.,  eds,  Astrophysics and Space Science Library Vol.
  407, Magnetic Fields in Diffuse Media. p.~599,
  \mn@doi{10.1007/978-3-662-44625-6_21}

\bibitem[\protect\citeauthoryear{{Bruno} \& {Carbone}}{{Bruno} \&
  {Carbone}}{2013}]{Bruno2013_soloar_wind_turbulence}
{Bruno} R.,  {Carbone} V.,  2013, \mn@doi [Living Reviews in Solar Physics]
  {10.12942/lrsp-2013-2}, \href
  {https://ui.adsabs.harvard.edu/abs/2013LRSP...10....2B} {10, 2}

\bibitem[\protect\citeauthoryear{Burgers}{Burgers}{1948}]{Burgers1948}
Burgers J.,  1948, \mn@doi [Advances in Applied Mechanics]
  {http://dx.doi.org/10.1016/S0065-2156(08)70100-5}, 1, 171

\bibitem[\protect\citeauthoryear{{Chernoglazov}, {Ripperda}  \&
  {Philippov}}{{Chernoglazov} et~al.}{2021}]{Chernoglazov2021_alignment_SR_MHD}
{Chernoglazov} A.,  {Ripperda} B.,   {Philippov} A.,  2021, \mn@doi [The
  Astrophysical Journal Letters] {10.3847/2041-8213/ac3afa}, \href
  {https://ui.adsabs.harvard.edu/abs/2021ApJ...923L..13C} {923, L13}

\bibitem[\protect\citeauthoryear{{Cielo}, {Iapichino}, {Baruffa}, {Bugli}  \&
  {Federrath}}{{Cielo} et~al.}{2020}]{Cielo2020_code_scaling}
{Cielo} S.,  {Iapichino} L.,  {Baruffa} F.,  {Bugli} M.,   {Federrath} C.,
  2020, \mn@doi [arXiv e-prints] {10.48550/arXiv.2002.08161}, \href
  {https://ui.adsabs.harvard.edu/abs/2020arXiv200208161C} {p. arXiv:2002.08161}

\bibitem[\protect\citeauthoryear{{Comisso}, {Huang}, {Lingam}, {Hirvijoki}  \&
  {Bhattacharjee}}{{Comisso}
  et~al.}{2018}]{Comisso2018_MHD_turbulence_plasmoid_regime}
{Comisso} L.,  {Huang} Y.~M.,  {Lingam} M.,  {Hirvijoki} E.,   {Bhattacharjee}
  A.,  2018, \mn@doi [The Astrophysical Journal] {10.3847/1538-4357/aaac83},
  \href {https://ui.adsabs.harvard.edu/abs/2018ApJ...854..103C} {854, 103}

\bibitem[\protect\citeauthoryear{{Dobrowolny}, {Mangeney}  \&
  {Veltri}}{{Dobrowolny} et~al.}{1980}]{Dobrowolny1980_solar_wind_properties}
{Dobrowolny} M.,  {Mangeney} A.,   {Veltri} P.,  1980, Astronomy and
  Astrophysics, \href {https://ui.adsabs.harvard.edu/abs/1980A&A....83...26D}
  {83, 26}

\bibitem[\protect\citeauthoryear{{Dong}, {Wang}, {Huang}, {Comisso},
  {Sandstrom}  \& {Bhattacharjee}}{{Dong}
  et~al.}{2022}]{Dong2022_reconnection_mediated_cascade}
{Dong} C.,  {Wang} L.,  {Huang} Y.-M.,  {Comisso} L.,  {Sandstrom} T.~A.,
  {Bhattacharjee} A.,  2022, \mn@doi [Science Advances]
  {10.1126/sciadv.abn7627}, \href
  {https://ui.adsabs.harvard.edu/abs/2022SciA....8N7627D} {8, eabn7627}

\bibitem[\protect\citeauthoryear{{Dubey} et~al.,}{{Dubey}
  et~al.}{2008}]{Dubey2008}
{Dubey} A.,  et~al., 2008, in {Pogorelov} N.~V.,  {Audit} E.,   {Zank} G.~P.,
  eds,  Astronomical Society of the Pacific Conference Series Vol. 385,
  Numerical Modeling of Space Plasma Flows. p.~145

\bibitem[\protect\citeauthoryear{{Dunn}, {Bowen}, {Mallet}, {Badman}  \&
  {Bale}}{{Dunn} et~al.}{2023}]{Dunn2023_steep_solar_wind_spectra}
{Dunn} C.,  {Bowen} T.~A.,  {Mallet} A.,  {Badman} S.~T.,   {Bale} S.~D.,
  2023, \mn@doi [The Astrophysical Journal] {10.3847/1538-4357/ad03ef}, \href
  {https://ui.adsabs.harvard.edu/abs/2023ApJ...958...88D} {958, 88}

\bibitem[\protect\citeauthoryear{{Eswaran} \& {Pope}}{{Eswaran} \&
  {Pope}}{1988}]{Eswaran1988_forcing_numerical_scheme}
{Eswaran} V.,  {Pope} S.~B.,  1988, Computers and Fluids, \href
  {https://ui.adsabs.harvard.edu/abs/1988CF.....16..257E} {16, 257}

\bibitem[\protect\citeauthoryear{Federrath}{Federrath}{2013}]{Federrath2013_universality}
Federrath C.,  2013, \mn@doi [The Monthly Notices of The Royal Astronomical
  Society] {10.1093/mnras/stt1644}, 436, 1245

\bibitem[\protect\citeauthoryear{{Federrath}}{{Federrath}}{2016}]{Federrath2016_dynamo}
{Federrath} C.,  2016, \mn@doi [Journal of Plasma Physics]
  {10.1017/S0022377816001069}, \href
  {https://ui.adsabs.harvard.edu/abs/2016JPlPh..82f5301F} {82, 535820601}

\bibitem[\protect\citeauthoryear{Federrath \& Klessen}{Federrath \&
  Klessen}{2012}]{Federrath2012}
Federrath C.,  Klessen R.~S.,  2012, \mn@doi [The Astrophysical Journal]
  {10.1088/0004-637X/761/2/156}, 761

\bibitem[\protect\citeauthoryear{Federrath, Roman-Duval, Klessen, Schmidt  \&
  {Mac Low}}{Federrath
  et~al.}{2010}]{Federrath2010_solendoidal_versus_compressive}
Federrath C.,  Roman-Duval J.,  Klessen R.,  Schmidt W.,   {Mac Low} M.~M.,
  2010, \mn@doi [Astronomy and Astrophysics] {10.1051/0004-6361/200912437}, 512

\bibitem[\protect\citeauthoryear{{Federrath} et~al.,}{{Federrath}
  et~al.}{2016}]{Federrath2016_brick}
{Federrath} C.,  et~al., 2016, \mn@doi [The Astrophysical Journal]
  {10.3847/0004-637X/832/2/143}, \href
  {https://ui.adsabs.harvard.edu/\#abs/2016ApJ...832..143F} {832, 143}

\bibitem[\protect\citeauthoryear{Federrath, Klessen, Iapichino  \&
  Beattie}{Federrath et~al.}{2021}]{Federrath2021}
Federrath C.,  Klessen R.~S.,  Iapichino L.,   Beattie J.~R.,  2021, \mn@doi
  [Nature Astronomy] {10.1038/s41550-020-01282-z}

\bibitem[\protect\citeauthoryear{{Federrath}, {Roman-Duval}, {Klessen},
  {Schmidt}  \& {Mac Low}}{{Federrath}
  et~al.}{2022}]{Federrath2022_turbulence_driving_module}
{Federrath} C.,  {Roman-Duval} J.,  {Klessen} R.~S.,  {Schmidt} W.,   {Mac Low}
  M.~M.,  2022, {TG: Turbulence Generator}, Astrophysics Source Code Library,
  record ascl:2204.001 (\mn@eprint {ascl} {2204.001})

\bibitem[\protect\citeauthoryear{{Ferri{\`e}re}}{{Ferri{\`e}re}}{2020}]{Ferriere2020_reynolds_numbers_for_ism}
{Ferri{\`e}re} K.,  2020, \mn@doi [Plasma Physics and Controlled Fusion]
  {10.1088/1361-6587/ab49eb}, \href
  {https://ui.adsabs.harvard.edu/abs/2020PPCF...62a4014F} {62, 014014}

\bibitem[\protect\citeauthoryear{{Fielding}, {Ripperda}  \&
  {Philippov}}{{Fielding} et~al.}{2023}]{Fielding2022_ISM_plasmoids}
{Fielding} D.~B.,  {Ripperda} B.,   {Philippov} A.~A.,  2023, \mn@doi [The
  Astrophysical Journal Letters] {10.3847/2041-8213/accf1f}, \href
  {https://ui.adsabs.harvard.edu/abs/2023ApJ...949L...5F} {949, L5}

\bibitem[\protect\citeauthoryear{{Fryxell} et~al.,}{{Fryxell}
  et~al.}{2000}]{Fryxell2000}
{Fryxell} B.,  et~al., 2000, \mn@doi [The Astrophysical Journal Supplement]
  {10.1086/317361}, \href
  {https://ui.adsabs.harvard.edu/abs/2000ApJS..131..273F} {131, 273}

\bibitem[\protect\citeauthoryear{{Galishnikova}, {Kunz}  \&
  {Schekochihin}}{{Galishnikova}
  et~al.}{2022}]{Galishnikova2022_saturation_and_tearing}
{Galishnikova} A.~K.,  {Kunz} M.~W.,   {Schekochihin} A.~A.,  2022, \mn@doi
  [Physical Review X] {10.1103/PhysRevX.12.041027}, \href
  {https://ui.adsabs.harvard.edu/abs/2022PhRvX..12d1027G} {12, 041027}

\bibitem[\protect\citeauthoryear{{Goldreich} \& {Sridhar}}{{Goldreich} \&
  {Sridhar}}{1995}]{Goldreich1995}
{Goldreich} P.,  {Sridhar} S.,  1995, \mn@doi [The Astrophysical Journal]
  {10.1086/175121}, \href
  {https://ui.adsabs.harvard.edu/abs/1995ApJ...438..763G} {438, 763}

\bibitem[\protect\citeauthoryear{{Grete}, {O'Shea}, {Beckwith}, {Schmidt}  \&
  {Christlieb}}{{Grete} et~al.}{2017}]{Grete2017_shell_models_for_CMHD}
{Grete} P.,  {O'Shea} B.~W.,  {Beckwith} K.,  {Schmidt} W.,   {Christlieb} A.,
  2017, \mn@doi [Physics of Plasmas] {10.1063/1.4990613}, \href
  {https://ui.adsabs.harvard.edu/abs/2017PhPl...24i2311G} {24, 092311}

\bibitem[\protect\citeauthoryear{{Grete}, {O'Shea}  \& {Beckwith}}{{Grete}
  et~al.}{2020}]{Grete2020_as_a_matter_of_state}
{Grete} P.,  {O'Shea} B.~W.,   {Beckwith} K.,  2020, \mn@doi [The Astrophysical
  Journal] {10.3847/1538-4357/ab5aec}, \href
  {https://ui.adsabs.harvard.edu/abs/2020ApJ...889...19G} {889, 19}

\bibitem[\protect\citeauthoryear{{Grete}, {O'Shea}  \& {Beckwith}}{{Grete}
  et~al.}{2021}]{Grete2021_as_a_matter_of_tension}
{Grete} P.,  {O'Shea} B.~W.,   {Beckwith} K.,  2021, \mn@doi [The Astrophysical
  Journal] {10.3847/1538-4357/abdd22}, \href
  {https://ui.adsabs.harvard.edu/abs/2021ApJ...909..148G} {909, 148}

\bibitem[\protect\citeauthoryear{{Grete}, {O'Shea}  \& {Beckwith}}{{Grete}
  et~al.}{2023}]{Grete2023_as_a_matter_of_dynamical_range}
{Grete} P.,  {O'Shea} B.~W.,   {Beckwith} K.,  2023, \mn@doi [The Astrophysical
  Journal Letters] {10.3847/2041-8213/acaea7}, \href
  {https://ui.adsabs.harvard.edu/abs/2023ApJ...942L..34G} {942, L34}

\bibitem[\protect\citeauthoryear{{Hopkins}}{{Hopkins}}{2013}]{Hopkins2013_non_lognormal_s_pdf}
{Hopkins} P.~F.,  2013, \mn@doi [The Monthly Notices of The Royal Astronomical
  Society] {10.1093/mnras/stt010}, \href
  {https://ui.adsabs.harvard.edu/abs/2013MNRAS.430.1880H} {430, 1880}

\bibitem[\protect\citeauthoryear{{Hopkins} et~al.,}{{Hopkins}
  et~al.}{2024}]{Hopkins2023_forged_in_fire_II}
{Hopkins} P.~F.,  et~al., 2024, \mn@doi [The Open Journal of Astrophysics]
  {10.21105/astro.2310.04506}, \href
  {https://ui.adsabs.harvard.edu/abs/2024OJAp....7E..19H} {7, 19}

\bibitem[\protect\citeauthoryear{{Hosking}, {Schekochihin}  \&
  {Balbus}}{{Hosking} et~al.}{2020}]{Hosking2020_tangled_field_stats}
{Hosking} D.~N.,  {Schekochihin} A.~A.,   {Balbus} S.~A.,  2020, \mn@doi
  [Journal of Plasma Physics] {10.1017/S0022377820001191}, \href
  {https://ui.adsabs.harvard.edu/abs/2020JPlPh..86e9011H} {86, 905860511}

\bibitem[\protect\citeauthoryear{{Iroshnikov}}{{Iroshnikov}}{1964}]{Iroshnikov_1965_IK_turb}
{Iroshnikov} P.~S.,  1964, Soviet Astronomy, \href
  {https://ui.adsabs.harvard.edu/abs/1964SvA.....7..566I} {7, 566}

\bibitem[\protect\citeauthoryear{{Kazantsev}}{{Kazantsev}}{1968}]{Kazantsev1968}
{Kazantsev} A.~P.,  1968, Soviet Journal of Experimental and Theoretical
  Physics, \href {https://ui.adsabs.harvard.edu/abs/1968JETP...26.1031K} {26,
  1031}

\bibitem[\protect\citeauthoryear{{Kempski} \& {Quataert}}{{Kempski} \&
  {Quataert}}{2022}]{Kempski2022_cr_scattering}
{Kempski} P.,  {Quataert} E.,  2022, \mn@doi [The Monthly Notices of The Royal
  Astronomical Society] {10.1093/mnras/stac1240}, \href
  {https://ui.adsabs.harvard.edu/abs/2022MNRAS.514..657K} {514, 657}

\bibitem[\protect\citeauthoryear{Kolmogorov}{Kolmogorov}{1941}]{Kolmogorov1941}
Kolmogorov A.~N.,  1941, \mn@doi [Doklady Akademii Nauk Sssr]
  {10.1098/rspa.1991.0075}, 30, 301

\bibitem[\protect\citeauthoryear{Kraichnan}{Kraichnan}{1965}]{Kraichnan1965_IKturb}
Kraichnan R.~H.,  1965, \mn@doi [The Physics of Fluids] {10.1063/1.1761412}, 8,
  1385

\bibitem[\protect\citeauthoryear{{Kriel}, {Beattie}, {Seta}  \&
  {Federrath}}{{Kriel} et~al.}{2022}]{Kriel2022_kinematic_dynamo_scales}
{Kriel} N.,  {Beattie} J.~R.,  {Seta} A.,   {Federrath} C.,  2022, \mn@doi [The
  Monthly Notices of The Royal Astronomical Society] {10.1093/mnras/stac969},
  \href {https://ui.adsabs.harvard.edu/abs/2022MNRAS.513.2457K} {513, 2457}

\bibitem[\protect\citeauthoryear{Kritsuk, Norman, Padoan  \& Wagner}{Kritsuk
  et~al.}{2007}]{Kritsuk2007}
Kritsuk A.~G.,  Norman M.~L.,  Padoan P.,   Wagner R.,  2007, \mn@doi [The
  Astrophysical Journal] {10.1086/519443}, 665, 416

\bibitem[\protect\citeauthoryear{{Lithwick} \& {Goldreich}}{{Lithwick} \&
  {Goldreich}}{2001}]{Lithwick2001_compressibleMHD}
{Lithwick} Y.,  {Goldreich} P.,  2001, \mn@doi [The Astrophysical Journal]
  {10.1086/323470}, \href
  {https://ui.adsabs.harvard.edu/abs/2001ApJ...562..279L} {562, 279}

\bibitem[\protect\citeauthoryear{{Loureiro} \& {Boldyrev}}{{Loureiro} \&
  {Boldyrev}}{2017}]{Loureiro2017_reconnection_in_turbulence}
{Loureiro} N.~F.,  {Boldyrev} S.,  2017, \mn@doi [Physical Review Letters]
  {10.1103/PhysRevLett.118.245101}, \href
  {https://ui.adsabs.harvard.edu/abs/2017PhRvL.118x5101L} {118, 245101}

\bibitem[\protect\citeauthoryear{{Loureiro} \& {Boldyrev}}{{Loureiro} \&
  {Boldyrev}}{2020}]{Loureiro2020_reconnection_in_turbulence}
{Loureiro} N.~F.,  {Boldyrev} S.,  2020, \mn@doi [The Astrophysical Journal]
  {10.3847/1538-4357/ab6a95}, \href
  {https://ui.adsabs.harvard.edu/abs/2020ApJ...890...55L} {890, 55}

\bibitem[\protect\citeauthoryear{Loureiro \& Uzdensky}{Loureiro \&
  Uzdensky}{2015}]{Loureiro_2016_plasmoid_instability}
Loureiro N.~F.,  Uzdensky D.~A.,  2015, \mn@doi [Plasma Physics and Controlled
  Fusion] {10.1088/0741-3335/58/1/014021}, 58, 014021

\bibitem[\protect\citeauthoryear{Mac~Low \& Klessen}{Mac~Low \&
  Klessen}{2004}]{MacLow2004}
Mac~Low M.~M.,  Klessen R.~S.,  2004, \mn@doi [Reviews of Modern Physics]
  {10.1103/RevModPhys.76.125}, 76, 125

\bibitem[\protect\citeauthoryear{{Mallet} \& {Schekochihin}}{{Mallet} \&
  {Schekochihin}}{2017}]{Mallet2017_anisotropy}
{Mallet} A.,  {Schekochihin} A.~A.,  2017, \mn@doi [The Monthly Notices of The
  Royal Astronomical Society] {10.1093/mnras/stw3251}, \href
  {https://ui.adsabs.harvard.edu/abs/2017MNRAS.466.3918M} {466, 3918}

\bibitem[\protect\citeauthoryear{{Mallet}, {Schekochihin}  \&
  {Chandran}}{{Mallet} et~al.}{2015}]{Mallet2015_refined_cb}
{Mallet} A.,  {Schekochihin} A.~A.,   {Chandran} B.~D.~G.,  2015, \mn@doi [The
  Monthly Notices of The Royal Astronomical Society] {10.1093/mnrasl/slv021},
  \href {https://ui.adsabs.harvard.edu/abs/2015MNRAS.449L..77M} {449, L77}

\bibitem[\protect\citeauthoryear{{Mallet}, {Schekochihin}  \&
  {Chandran}}{{Mallet} et~al.}{2017}]{Mallet2017_plasmoid_disruptions}
{Mallet} A.,  {Schekochihin} A.~A.,   {Chandran} B.~D.~G.,  2017, \mn@doi [The
  Monthly Notices of The Royal Astronomical Society] {10.1093/mnras/stx670},
  \href {https://ui.adsabs.harvard.edu/abs/2017MNRAS.468.4862M} {468, 4862}

\bibitem[\protect\citeauthoryear{{Malvadi Shivakumar} \& {Federrath}}{{Malvadi
  Shivakumar} \& {Federrath}}{2023}]{Shivakumar2023_numerical_dissipation}
{Malvadi Shivakumar} L.,  {Federrath} C.,  2023, \mn@doi [arXiv e-prints]
  {10.48550/arXiv.2311.10350}, \href
  {https://ui.adsabs.harvard.edu/abs/2023arXiv231110350M} {p. arXiv:2311.10350}

\bibitem[\protect\citeauthoryear{{Maron} \& {Goldreich}}{{Maron} \&
  {Goldreich}}{2001}]{Maron2001}
{Maron} J.,  {Goldreich} P.,  2001, \mn@doi [The Astrophysical Journal]
  {10.1086/321413}, \href
  {https://ui.adsabs.harvard.edu/abs/2001ApJ...554.1175M} {554, 1175}

\bibitem[\protect\citeauthoryear{{Matthaeus}, {Pouquet}, {Mininni}, {Dmitruk}
  \& {Breech}}{{Matthaeus} et~al.}{2008}]{Matthaeus2008_rapid_alignment}
{Matthaeus} W.~H.,  {Pouquet} A.,  {Mininni} P.~D.,  {Dmitruk} P.,   {Breech}
  B.,  2008, \mn@doi [Physical Review Letters]
  {10.1103/PhysRevLett.100.085003}, \href
  {https://ui.adsabs.harvard.edu/abs/2008PhRvL.100h5003M} {100, 085003}

\bibitem[\protect\citeauthoryear{{Meyrand}, {Galtier}  \& {Kiyani}}{{Meyrand}
  et~al.}{2016}]{Meyrand2016_strong_to_weak_transition}
{Meyrand} R.,  {Galtier} S.,   {Kiyani} K.~H.,  2016, \mn@doi [Physical Review
  Letters] {10.1103/PhysRevLett.116.105002}, \href
  {https://ui.adsabs.harvard.edu/abs/2016PhRvL.116j5002M} {116, 105002}

\bibitem[\protect\citeauthoryear{{Mocz} \& {Burkhart}}{{Mocz} \&
  {Burkhart}}{2018}]{Mocz2018}
{Mocz} P.,  {Burkhart} B.,  2018, \mn@doi [The Monthly Notices of The Royal
  Astronomical Society] {10.1093/mnras/sty1976}, \href
  {https://ui.adsabs.harvard.edu/\#abs/2018MNRAS.480.3916M} {480, 3916}

\bibitem[\protect\citeauthoryear{Mocz \& Burkhart}{Mocz \&
  Burkhart}{2019}]{Mocz2019}
Mocz P.,  Burkhart B.,  2019, \mn@doi [The Astrophysical Journal Letters]
  {10.3847/2041-8213/ab48f6}, 884, L35

\bibitem[\protect\citeauthoryear{{Mohapatra} \& {Sharma}}{{Mohapatra} \&
  {Sharma}}{2019}]{Mohapatra2019_turbulent_heat_flux_ICM}
{Mohapatra} R.,  {Sharma} P.,  2019, \mn@doi [The Monthly Notices of The Royal
  Astronomical Society] {10.1093/mnras/stz328}, \href
  {https://ui.adsabs.harvard.edu/abs/2019MNRAS.484.4881M} {484, 4881}

\bibitem[\protect\citeauthoryear{M\"uller \& Grappin}{M\"uller \&
  Grappin}{2005}]{Muller2005_k32_total_spectrum}
M\"uller W.-C.,  Grappin R.,  2005, \mn@doi [Physical Review Letters]
  {10.1103/PhysRevLett.95.114502}, 95, 114502

\bibitem[\protect\citeauthoryear{{Osman}, {Wan}, {Matthaeus}, {Breech}  \&
  {Oughton}}{{Osman} et~al.}{2011}]{Osman2011_solar_wind_alignment}
{Osman} K.~T.,  {Wan} M.,  {Matthaeus} W.~H.,  {Breech} B.,   {Oughton} S.,
  2011, \mn@doi [The Astrophysical Journal] {10.1088/0004-637X/741/2/75}, \href
  {https://ui.adsabs.harvard.edu/abs/2011ApJ...741...75O} {741, 75}

\bibitem[\protect\citeauthoryear{{Pecora} et~al.,}{{Pecora}
  et~al.}{2023}]{Pecora2023_relaxation_in_magnetosheath}
{Pecora} F.,  et~al., 2023, \mn@doi [The Monthly Notices of The Royal
  Astronomical Society] {10.1093/mnras/stad2232}, \href
  {https://ui.adsabs.harvard.edu/abs/2023MNRAS.525...67P} {525, 67}

\bibitem[\protect\citeauthoryear{{Perez} \& {Boldyrev}}{{Perez} \&
  {Boldyrev}}{2008}]{Perez2008_weak_and_strong_turb}
{Perez} J.~C.,  {Boldyrev} S.,  2008, \mn@doi [The Astrophysical Journal
  Letters] {10.1086/526342}, \href
  {https://ui.adsabs.harvard.edu/abs/2008ApJ...672L..61P} {672, L61}

\bibitem[\protect\citeauthoryear{{Perez} \& {Boldyrev}}{{Perez} \&
  {Boldyrev}}{2009}]{Perez2009_dynamical_alignment_of_imbalanced_islands}
{Perez} J.~C.,  {Boldyrev} S.,  2009, \mn@doi [Physical Review Letters]
  {10.1103/PhysRevLett.102.025003}, \href
  {https://ui.adsabs.harvard.edu/abs/2009PhRvL.102b5003P} {102, 025003}

\bibitem[\protect\citeauthoryear{{Price} \& {Federrath}}{{Price} \&
  {Federrath}}{2010}]{Price2010_grid_versus_SPH}
{Price} D.~J.,  {Federrath} C.,  2010, \mn@doi [The Monthly Notices of The
  Royal Astronomical Society] {10.1111/j.1365-2966.2010.16810.x}, \href
  {https://ui.adsabs.harvard.edu/abs/2010MNRAS.406.1659P} {406, 1659}

\bibitem[\protect\citeauthoryear{{Rincon}}{{Rincon}}{2019}]{Rincon2019_dynamo_theories}
{Rincon} F.,  2019, \mn@doi [Journal of Plasma Physics]
  {10.1017/S0022377819000539}, \href
  {https://ui.adsabs.harvard.edu/abs/2019JPlPh..85d2001R} {85, 205850401}

\bibitem[\protect\citeauthoryear{{Robertson} \& {Goldreich}}{{Robertson} \&
  {Goldreich}}{2018}]{Robertson2018}
{Robertson} B.,  {Goldreich} P.,  2018, \mn@doi [The Astrophysical Journal]
  {10.3847/1538-4357/aaa89e}, \href
  {https://ui.adsabs.harvard.edu/abs/2018ApJ...854...88R} {854, 88}

\bibitem[\protect\citeauthoryear{{Rosotti}}{{Rosotti}}{2023}]{Rosotti2023_supersonic_turbulence_in_pp_discs}
{Rosotti} G.~P.,  2023, \mn@doi [New Astronomy Reviews]
  {10.1016/j.newar.2023.101674}, \href
  {https://ui.adsabs.harvard.edu/abs/2023NewAR..9601674R} {96, 101674}

\bibitem[\protect\citeauthoryear{{Schekochihin}}{{Schekochihin}}{2022}]{Schekochihin2020_bias_review}
{Schekochihin} A.~A.,  2022, \mn@doi [Journal of Plasma Physics]
  {10.1017/S0022377822000721}, \href
  {https://ui.adsabs.harvard.edu/abs/2022JPlPh..88e1501S} {88, 155880501}

\bibitem[\protect\citeauthoryear{Schekochihin, Cowley, Hammett, Maron  \&
  McWilliams}{Schekochihin
  et~al.}{2002}]{Schekochihin2002_saturation_evolution}
Schekochihin A.~A.,  Cowley S.~C.,  Hammett G.~W.,  Maron J.~L.,   McWilliams
  J.~C.,  2002, \mn@doi [New Journal of Physics] {10.1088/1367-2630/4/1/384},
  4, 84

\bibitem[\protect\citeauthoryear{Schekochihin, Cowley, Taylor, Maron  \&
  McWilliams}{Schekochihin et~al.}{2004}]{Schekochihin2004_dynamo}
Schekochihin A.~A.,  Cowley S.~C.,  Taylor S.~F.,  Maron J.~L.,   McWilliams
  J.~C.,  2004, \mn@doi [The Astrophysical Journal] {10.1086/422547}, 612, 276

\bibitem[\protect\citeauthoryear{{Schmidt}, {Federrath}, {Hupp}, {Kern}  \&
  {Niemeyer}}{{Schmidt} et~al.}{2009}]{Schmidt2009}
{Schmidt} W.,  {Federrath} C.,  {Hupp} M.,  {Kern} S.,   {Niemeyer} J.~C.,
  2009, \mn@doi [Astronomy and Astrophysics] {10.1051/0004-6361:200809967},
  \href {https://ui.adsabs.harvard.edu/\#abs/2009A&A...494..127S} {494, 127}

\bibitem[\protect\citeauthoryear{{Schober}, {Schleicher}, {Federrath}, {Bovino}
   \& {Klessen}}{{Schober}
  et~al.}{2015}]{Schober2015_saturation_of_turbulent_dynamo}
{Schober} J.,  {Schleicher} D.~R.~G.,  {Federrath} C.,  {Bovino} S.,
  {Klessen} R.~S.,  2015, \mn@doi [Physical Review E]
  {10.1103/PhysRevE.92.023010}, \href
  {https://ui.adsabs.harvard.edu/abs/2015PhRvE..92b3010S} {92, 023010}

\bibitem[\protect\citeauthoryear{{Seta} \& {Federrath}}{{Seta} \&
  {Federrath}}{2020}]{Seta2020_seed_magnetic_field}
{Seta} A.,  {Federrath} C.,  2020, \mn@doi [The Monthly Notices of The Royal
  Astronomical Society] {10.1093/mnras/staa2978}, \href
  {https://ui.adsabs.harvard.edu/abs/2020MNRAS.499.2076S} {499, 2076}

\bibitem[\protect\citeauthoryear{{Seta} \& {Federrath}}{{Seta} \&
  {Federrath}}{2021}]{Seta2021_supersonic_saturation}
{Seta} A.,  {Federrath} C.,  2021, \mn@doi [Physical Review Fluids]
  {10.1103/PhysRevFluids.6.103701}, \href
  {https://ui.adsabs.harvard.edu/abs/2021PhRvF...6j3701S} {6, 103701}

\bibitem[\protect\citeauthoryear{She \& Leveque}{She \&
  Leveque}{1994}]{She1994}
She Z.-S.,  Leveque E.,  1994, \mn@doi [Physical Review Letters]
  {10.1103/PhysRevLett.72.336}, 72, 336

\bibitem[\protect\citeauthoryear{{St-Onge}, {Kunz}, {Squire}  \&
  {Schekochihin}}{{St-Onge}
  et~al.}{2020}]{StOnge2020_weakly_collisional_dynamo}
{St-Onge} D.~A.,  {Kunz} M.~W.,  {Squire} J.,   {Schekochihin} A.~A.,  2020,
  \mn@doi [Journal of Plasma Physics] {10.1017/S0022377820000860}, \href
  {https://ui.adsabs.harvard.edu/abs/2020JPlPh..86e9003S} {86, 905860503}

\bibitem[\protect\citeauthoryear{Stribling \& Matthaeus}{Stribling \&
  Matthaeus}{1991}]{Stribling1991_relaxation_processes}
Stribling T.,  Matthaeus W.~H.,  1991, \mn@doi [Physics of Fluids B: Plasma
  Physics] {10.1063/1.859654}, 3, 1848

\bibitem[\protect\citeauthoryear{Uzdensky, Loureiro  \& Schekochihin}{Uzdensky
  et~al.}{2010}]{Uzdensky2010_fast_reconnection}
Uzdensky D.~A.,  Loureiro N.~F.,   Schekochihin A.~A.,  2010, \mn@doi [Physical
  Review Letters] {10.1103/PhysRevLett.105.235002}, 105, 235002

\bibitem[\protect\citeauthoryear{{Waagan}, {Federrath}  \&
  {Klingenberg}}{{Waagan} et~al.}{2011}]{Waagan2011}
{Waagan} K.,  {Federrath} C.,   {Klingenberg} C.,  2011, \mn@doi [Journal of
  Computational Physics] {10.1016/j.jcp.2011.01.026}, \href
  {https://ui.adsabs.harvard.edu/abs/2011JCoPh.230.3331W} {230, 3331}

\bibitem[\protect\citeauthoryear{{Xu} \& {Lazarian}}{{Xu} \&
  {Lazarian}}{2016}]{Xu2016_dynamo}
{Xu} S.,  {Lazarian} A.,  2016, \mn@doi [The Astrophysical Journal]
  {10.3847/1538-4357/833/2/215}, \href
  {https://ui.adsabs.harvard.edu/abs/2016ApJ...833..215X} {833, 215}

\bibitem[\protect\citeauthoryear{Zank \& Matthaeus}{Zank \&
  Matthaeus}{1992}]{Zank_1992_waves_in_solar_wind}
Zank G.~P.,  Matthaeus W.~H.,  1992, \mn@doi [Journal of Geophysical Research:
  Space Physics] {https://doi.org/10.1029/92JA01734}, 97, 17189

\makeatother
\end{thebibliography}
